\documentclass[preprint]{elsarticle}
\RequirePackage{amsthm,amsmath,natbib}
\RequirePackage[colorlinks,citecolor=blue,urlcolor=blue]{hyperref}

\usepackage{epsfig}
\usepackage{latexsym,amsfonts,amsmath,amssymb,amsthm}
\usepackage[active]{srcltx}
\usepackage{graphicx}
\usepackage{psfrag}
\usepackage[active]{srcltx}
\usepackage{amsmath,amsfonts,amssymb,latexsym,epsfig}
\usepackage{paralist}
\usepackage{array}

\newtheorem{thm}{Theorem}[section]
\newtheorem{ass}{Assumption}[section]
\newtheorem{prop}{Proposition}[section]

\newtheorem{rem}{Remark}[section]
\newcommand{\R}{\mathbb{R}}
\newcommand{\Cov}{\mathcal{C}}
\newcommand{\Exp}{\mathrm{E}}
\newcommand{\psih}{\psi_{h}}
\newcommand{\Hs}{\mathcal{H}}
\newcommand{\psiT}{\psi_{h}^{I}}
\newcommand{\PsiT}{\Psi_{h}^{I}}

\newcommand{\XiL}{\tilde{\Xi}_{h^{\ast}}}
\newcommand{\PsiInv}[1]{\Psi_{h}^{-#1}}

\begin{document}

\begin{frontmatter}

\title{Advanced MCMC Methods for Sampling on Diffusion Pathspace}


\author{Alexandros Beskos}
\address{Department of Statistical Science, University College London, UK}
\ead{a.beskos@ucl.ac.uk}

\author{Konstantinos Kalogeropoulos}
\address{Department of Statistics, London School of Economics}
\ead{K.Kalogeropoulos@lse.ac.uk}

\author{Erik Pazos}
\address{Department of Statistical Science, University College London, UK}
\ead{erikpazos05@gmail.com}

\author{The paper is also available in the journal `Stochastic Processes and Their Applications'}

\begin{abstract}

The need to calibrate increasingly complex statistical models requires a persistent effort for further advances on available, computationally intensive Monte-Carlo methods. We study here an advanced version of familiar Markov-chain Monte-Carlo (MCMC) algorithms that sample from target distributions defined as change of measures from Gaussian laws on general Hilbert spaces. Such a model structure arises in several contexts: we focus here at the important  class of statistical models driven by diffusion paths whence the Wiener process constitutes the reference Gaussian law. Particular emphasis is given on advanced Hybrid Monte-Carlo (HMC) which makes large, derivative-driven steps in the state space (in contrast with local-move Random-walk-type algorithms) with analytical and experimental results. We illustrate its computational advantages in various diffusion processes and observation regimes; examples include stochastic volatility and latent survival models. In contrast with their standard MCMC counterparts, the advanced versions have \emph{mesh-free} mixing times, as these will not deteriorate upon refinement of the approximation of the inherently infinite-dimensional diffusion paths by finite-dimensional ones used in practice when applying the algorithms on a computer.
\end{abstract}


\end{frontmatter}

%
%








\section{Introduction}

Markov chain Monte-Carlo (MCMC) methods provide an intuitive, powerful mechanism for sampling from complex posterior distributions arising in applications (see \cite{gilk:96} for a review of algorithms and applications). The rapidly increasing complexity of statistical models employed by practitioners requires a parallel effort at advancing MCMC methodology to deliver algorithms that can provide fast model exploration. Ideally, suggested algorithms should be of enough flexibility to cover a wide range of model structures. This paper will move along such directions by proposing and studying advanced versions of standard MCMC algorithms to improve algorithmic performance on complex, high-dimensional models.

The advanced algorithms are relevant for target distributions defined as change of measures from Gaussian laws. Within such a structure, we will be focusing here upon the important class of statistical models driven by Stochastic Differential Equations (SDEs) posing concrete computational challenges; here, Brownian motion constitutes the reference Gaussian measure. The paper will develop and test advanced MCMC algorithms for the computationally demanding task of reproducing sample paths of SDEs under various direct or indirect observation regimes. The ability to sample the realized dynamics driving the data mechanism is of high importance for understanding the model behavior. Also, due to the typical intractability of the likelihood function in SDE contexts, the underlying diffusion path is many times treated as a latent variable within Gibbs samplers and its fast sampling is critical for the efficiency and feasibility of parametric inference procedures.

The advanced MCMC methods follow closely recent developments \cite{besk:08, besk:11} over algorithms that take advantage of
the structure of target distributions $\Pi$ being determined as a change of measure from a Gaussian one $\Pi_0\equiv N(0,\Cov)$, that is:
\begin{equation}
\label{eq:target}
\frac{d	\Pi}{d\Pi_0}(x) = \exp\{-\Phi(x)\} \ ,
\end{equation}
for some function $\Phi$ defined on a Hilbert space $\Hs$. The method exploits the relation with the Gaussian measure to evolve standard MCMC algorithms into advanced ones with the critical computational advantage that their convergence properties are \emph{mesh-free}, i.e.\@ their mixing times do not deteriorate as the dimension of the state space increases when refining relevant finite-dimensional projections (used in practice on a computer) to better approximate inherently infinite-dimensional elements of~$\Hs$. In the SDE context, finite-difference methods are commonly employed to approximate the infinite-dimensional
diffusion sample paths whence a discretization mesh will be specified. We will be looking on advanced versions of Random-walk Metropolis (RWM), Metropolis-adjusted Langevin algorithm (MALA) and Hybrid Monte-Carlo (HMC). MALA and HMC both use information about the derivative of the log-target to drive the algorithmic dynamics, whereas RWM uses blind proposals. Emphasis will be given on HMC, employing Hamiltonian dynamics, as its nature to perform global designated steps on the SDE pathspace seems to
provide significant computational advantages compared to the other two local-move algorithms; this will be illustrated
both analytically and experimentally.

A first methodological contribution of the paper will be to define HMC on a Hilbert space, similar
to Hilbert-space valued RWM and MALA derived in \cite{besk:08}. A similar attempt for HMC has
 taken place in \cite{besk:11}; the method there required strict, \emph{analytical} Lipschitz-continuity
assumptions on the $\Phi$-derivative $\delta\Phi$ \footnote{the $\delta$-notation refers in the most general
setting to the Fr\'echet
 generalisation of differentiation of real-valued functions defined on general
Hilbert spaces; in our case
we will work with the associated (to the differential operator) element of the dual space,
so $\delta \Phi$ will be treated as an
element of the pathspace.}. In this paper,
a much simpler derivation will be proven under minimal \emph{probabilistic} assumptions on $\Phi$
relevant for a much wider range of practical applications that, for instance, could involve inherently
stochastic terms (e.g.\@ stochastic integrals) at the specification of $\Phi$. The concrete consequence
 of the mathematical verification of a well-defined algorithm on the infinite-dimensional Hilbert space
 $\Hs$ is an anticipated mesh-free mixing time for the practical algorithm that will run on some
$N$-dimensional projection (typically, finite-difference) of the pathspace.
Another methodological contribution of the paper will be that it will analytically
illustrate, in the context of directly observed diffusions, that `clever' use of information
on the derivative of the log-target (within HMC, compared to MALA) can remove orders of complexity
 from the computational costs of the MCMC algorithm. Albeit proven in a linear scenario, the result
will be later on empirically manifested in applications on realistic models. In addition to verifying
 the theoretical results, the empirical contributions of the paper relate with the implementation of
the pathspace algorithms on several practical problems including stochastic volatility and various
diffusion driven models. As demonstrated in relevant applications, HMC algorithms avoid the
 random-walk-type behaviour of the other algorithms and greatly outperform them, thus constituting
a powerful method for tackling high-dimensional path-sampling problems.

\subsection{Diffusion-Driven Models}
\label{sec:introdif}

Stochastic Differential Equations (SDEs) provide a powerful and flexible probabilistic structure for modeling phenomena in a multitude of disciplines: finance, biology, molecular dynamics, chemistry, survival analysis, epidemiology, just to name a few
(see \cite{kloe:92} for a review of applications). A large class of SDEs can be specified as follows:
\begin{equation}
\label{eq:diffusion}
dV_u = \mu(V_u;\theta)du + \sigma(V_u;\theta)dB_u \ ,
\end{equation}
for drift and diffusion coefficient
$\mu(\cdot;\theta):\R^{d}\mapsto \R^{d}$ and $\sigma(\cdot;\theta):\R^{d}\times\R^{d}\mapsto \R^{d}\times\R^{d}$
 respectively (of known functional form up to some unknown parameter $\theta\in\R^{p}$), and Brownian motion
$B_u$. Such a differential modeling structure is sometimes implied by physical laws
(e.g.\@ Langevin equation in physics) or is selected due to its adaptiveness by the
practitioner to represent time-evolving quantities.

The diffusion process defined by (\ref{eq:diffusion}) may be observed in a number of ways.
The case where data form a discrete skeleton of the diffusion path has been studied extensively
(see for example \cite{saha:08,bes:pap:rob:f06,dur:gal02,bib:jac:sor10}), but different
types of observation regimes often appear in applications. For example in PK/PD studies \citep{Oveetal05},
growth curve models \citep{Don:etal10} and bioinformatics \citep{goli:08},
the stochastic dynamical systems are observed with error. In financial applications,
and in particular stochastic volatility models \citep{ghy:har:ren96},
the observation model consists of partial data or integrals of the underlying diffusion process
\citep{ch:pit:she06}. Event time models as in \cite{aal:gje04,rob:san10}, involve a
latent diffusion process that is observed through random barrier hitting times.
In this paper we demonstrate how the above framework can be unified through formulations
that consist of a diffusion process defined by (\ref{eq:diffusion}) and are completed by
an appropriate (typically Lebesgue) density $p(y|v,\theta)$ for the observations $Y$
conditionally on the diffusion path $V$ and $\theta$. We then proceed and develop
a general and efficient HMC algorithm to handle models within this framework.

Inferential procedures about $\theta$ are perplexed by the typical unavailability of the
likelihood function $p(y|\theta)$ in closed form. MCMC methods could in principle overcome
 such an issue via \emph{data augmentation}, whereby Gibbs sampler steps within the algorithm
 would switch from sampling the unobserved diffusion path $V$ conditionally on $\theta$ and $Y$
to sampling $\theta$ conditionally on $V$ and $Y$. Our advanced MCMC methods are relevant here
for the challenging high-dimensional path-update of $V$.

The construction of such data augmentation MCMC samplers has to address two important
issues arising from inherent characteristics of diffusion processes. The first issue is
the singularity between the parameters in the diffusion coefficient and the latent
diffusion paths (i.e.\@ conditionally on $V$, the parameters in the diffusion coefficient
have a Dirac distribution) caused by the quadratic variation process.
The problem was identified in \cite{rob:str01} noting that the convergence of an
MCMC Independence sampler deteriorates as $\mathcal{O}(N)$ unless appropriate
reparametrisation is applied ($N$ being the number of discrete instances of the
path considered in practice when executing the method on a personal computer).
Various such reparametrizations are currently available and some of them will be used in the sequel.
The second issue is the construction of efficient proposals for the update of
the (high-dimensional) latent paths. The efficiency and tuning of such updates
are critical for the overall performance of the algorithms. Most existing approaches
adopt Independence sampler steps proposing paths from easier-to-handle diffusion processes
(typically, Brownian motion paths) as candidates for paths of the target $V|Y,\theta$.
While these approaches may work well in some cases, they can also perform unacceptably
poorly in many other ones when target paths are very different from Brownian ones.

The advanced MCMC samplers studied in this paper connect with both above issues.
First, methods used for decoupling the dependence between $V$ and $\theta$,
achieve this via the consideration of some 1-1 transform $X=\eta(V;\theta)$ so that
the distribution of $X$ has a density w.r.t.\@ a probability measure that will not
involve $\theta$: this reference measure is precisely a Brownian motion (or a Brownian bridge).
Thus, the decoupling approaches conveniently deliver target distributions which are
change of measures from Gaussian measures of the structure (\ref{eq:target}).
More analytically, using the reparametrization method of \cite{goli:08} our advanced
samplers will be, in principle, applicable under the condition:
\begin{equation}
\label{eq:coef}
v\mapsto \sigma(v;\theta)\,\,\, \textrm{is invertible, for any $v$ in the state space of $V$\ .}
\end{equation}
For the second issue, the advanced HMC sampler provides a powerful mechanism for
delivering efficient updates for the high-dimensional latent paths as: (i) it will
 propose non-blind path updates, driven by the log-derivative of the target distribution;
(ii) proposed updates will make large steps in the path space thus offering the potential for efficient mixing;
(iii) it's mixing time will be mesh-free, i.e.\@ will not deteriorate with increasing $N$.

The paper is organised as follows: Section \ref{sec:gaussian} contains some background material regarding Gaussian distributions on separable Hilbert Spaces. The advanced HMC sampler is developed in Section \ref{sec:advHMC} where other MCMC samplers for diffusion pathspaces are also presented. In Section \ref{sec:OUanalytical} we demonstrate the advantages offered by the HMC algorithm through an analytical study on the Ornstein-Uhlenbeck (OU) diffusion process. The theory of Section \ref{sec:advHMC} is used in Section \ref{sec:ReducibleDiff} to extend the framework the HMC algorithm to a rich class family of diffusion models with general diffusion coefficients and various observation regimes. Section \ref{sec:Numerical} verifies the theoretical results and illustrates the developed algorithms in various simulation experiments. Further generalizations are provided in Section \ref{sec:GeneralDiff}, while Section \ref{sec:Discussion} concludes with some relevant discussion and future directions.

\section{Gaussian Measures on Hilbert Spaces}
\label{sec:gaussian}

We collect here some background material (see e.g.\@ \cite{prat:92}) on Gaussian distributions on a separable Hilbert space $\Hs$
that will assist at the presentation of the later sections.
The Cameron-Martin space, $\Hs_0$, of the Gaussian law $\Pi_0\equiv N(0,\Cov)$ coincides with the image space of $\Cov^{1/2}$.
Essentially, $\Hs_0$ includes all elements of the Hilbert space which preserve the absolute continuity properties of $\Pi_0$
upon translation. This is made mathematically explicit via the following proposition.

\begin{prop}
\label{prop:tran}
If $T(v) = v + \Cov^{1/2}v_0$ for a constant $v_0\in\Hs$ then $\Pi_0$ and $\Pi_0\circ T^{-1}$
are absolutely continuous w.r.t.\@ each other with density:
\begin{equation*}
\frac{d\,\{\,\Pi_0\circ T^{-1}\,\}}{d\Pi_0}(v) = \exp\big\{ \langle v_0, \Cov^{-1/2}v \rangle - \tfrac{1}{2}
  |v_0|^2 \big\}\ .
\end{equation*}
\end{prop}
\begin{proof}
This is Theorem 2.21 of  \cite{prat:92}.
\end{proof}

For the diffusion pathspace we  focus upon in this paper,  the target  distribution~$\Pi(dx)$
 is defined on the Hilbert space of squared integrable paths
$\Hs = L^{2}([0,\ell],\R)$ (with appropriate boundary conditions) for some length $\ell>0$.
The centered Gaussian reference measure $\Pi_0$ will correspond to a Brownian motion (thus, boundary condition $x(0)=0$) or a Brownian Bridge ($x(0)=x(\ell)=0$).
The covariance operator is connected with the covariance function $c(u,v)$ of the Gaussian process
via:
\begin{equation*}
(\Cov f)(u) = \int_{0}^{\ell} c(u,v)f(v) dv \ ,\quad f\in \Hs\ .
\end{equation*}
With this in mind, the covariance operators $\Cov^{bm}$, $\Cov^{bb}$ of the Brownian motion and Brownian bridge respectively are
as below:
\begin{align}
(\Cov^{bm}f)(u) &= \int_{0}^{\ell} (u\wedge v)\,f(v)dv  = u\int_{0}^{\ell}f(v)dv - \int_{0}^{u}\int_{0}^{s}f(v)dv\,ds \  \label{eq:bm};\\
(\Cov^{bb}f)(u) &= \int_{0}^{\ell} (u\wedge v-\tfrac{uv}{\ell})\,f(v)dv \nonumber \\ &= \frac{u}{\ell}\,\int_{0}^{\ell}\int_{0}^{s}f(v)dv\,ds -
\int_{0}^{u}\int_{0}^{s}f(v)dv\,ds \ .  \label{eq:bb}
\end{align}
The Cameron-Martin spaces $\Hs_0^{bm}$ and $\Hs_0^{bb}$ of a Brownian motion and Brownian bridge respectively
are analytically specified as follows (see e.g.\@ Lemma 2.3.14 of \cite{boga:98} for the case of Brownian motion;
Brownian bridge
involves the extra boundary condition $x(\ell)=0$):
\begin{align*}
\Hs^{bm}_0 &= \big\{
x:[0,\ell]\mapsto \R: \exists\,\, f\in L^{2}([0,\ell],\R)\textrm{ such that } x(u) = \int_{[0,u]}f(s)ds
\big\}\ ; \\
\Hs^{bb}_0 &= \big\{
x:[0,\ell]\mapsto \R: \exists\,\, f\in L^{2}([0,\ell],\R)\textrm{ such that } \\ & \qquad \qquad\qquad \qquad\qquad \qquad\qquad\qquad\qquad
 x(u) = \int_{[0,u]}f(s)ds,\,x(\ell)=0
\big\}\ .
\end{align*}

The Karhunen-Lo\`eve representation of $N(0,\Cov)$ will be used later on.
Analytically, considering the eigen-decomposition $\{\lambda_p,\phi_p\}_{p=1}^{\infty}$ of $\Cov$ so that
 $\Cov\, \phi_p = \lambda_p\,\phi_p $, we have that the (normalized) eigenfunctions $\{\phi_p\}_{p=1}^{\infty}$ constitute
an orthonormal basis for the Hilbert space $\Hs$. In particular, for $x\sim N(0,\Cov)$
we have the expansion:
\begin{equation}
\label{eq:KL-exp}
x=\sum_{p=1}^{\infty}\langle x,\phi_p \rangle \,\phi_p=
\sum_{p=1}^{\infty}x_p\,\phi_p=
\sum_{p=1}^{\infty} \sqrt{\lambda_p}\,\xi_p\,\phi_p\ ,
\end{equation}
where $\{\xi_p\}_{p=1}^{\infty}$ are iid variables from $N(0,1)$.


\section{Advanced HMC on Hilbert Space}
\label{sec:advHMC}

HMC uses the derivative of the log-target density, in the form of Hamiltonian dynamics, to generate moves on the state space.
An appropriate accept/reject decision will then
force reversibility w.r.t.\@ the target distribution.
Uniquely amongst alternative MCMC algorithms, HMC allows for the synthesis of a number
of derivative-guided steps before the accept/reject test, thus permitting large, designated steps
in the state space.

The standard version of HMC  first appeared in \cite{duna:87}.
An advanced version for infinite-dimensional targets of the form (\ref{eq:target})
appeared in \cite{besk:11}. One of our objectives here is to extend the relevance of
the method in \cite{besk:11} to a much wider class of SDE-driven applications.
In particular, \cite{besk:11} defines an HMC algorithm on a Hilbert space under
strong analytical Lipschitz-continuity assumptions on Sobolev norms of $\delta \Phi$ which are difficult to verify
and are, indeed, not relevant for the common scenario when $\Phi$ will involve inherently stochastic terms (like stochastic integrals).
Here, we provide an alternative derivation of advanced HMC which avoids such strong assumptions of earlier
works, and is relevant for many practical  applications.
Such a construction of the algorithm on the Hilbert space will provide a justification for order $\mathcal{O}(1)$-mixing times for
 the related $N$-dimensional projected algorithm. This justification will not be analytically proven here as it would require a level
of mathematical technicalities far beyond the application-motivated scope of this paper; we note that all numerical applications shown later will empirically
verify the  $\mathcal{O}(1)$-mixing times of the samplers.

We will only require the following condition (recall that $\Cov$ is the covariance operator of the Gaussian measure
at the definition of the target $\Pi$ in (\ref{eq:target})):

\begin{ass}
\label{ass:ass}
$\Cov\,\delta \Phi(x)$ is an element of the Cameron-Martin space of the Gaussian measure
$\Pi_0$ (so\, $\Cov\,\delta \Phi(x)\in \mathrm{Im}\,\Cov^{1/2}$) for all $x$ in a set
with probability 1 under $\Pi_0$.
\end{ass}

We will discuss Assumption \ref{ass:ass} in the context of our applications in the sequel.
Note that the final algorithm suggested later in this section coincides with the one presented in \cite{besk:11} when the strict analytical conditions of that paper
on $\Phi$ are
satisfied; the new theoretical development here is to construct an alternative proof for the well-definition of the algorithm on
the Hilbert space that requires only Assumption \ref{ass:ass} and, as a consequence, greatly extends the relevance of the developed method for applications.

Some of the calculations carried out in the development of the algorithm in the sequel make explicit sense
for $\Hs=\R^{N}$ and  only formal sense when $\Hs$ is an infinite-dimensional separable Hilbert space.
We can write (formally in the case of infinite-dimensional $\Hs$):
\begin{equation}
\label{eq:targetR}
 \Pi(x) \propto \exp\{-\Phi(x)-\tfrac{1}{2}\langle x, Lx \rangle \}\ ,
\end{equation}
where we have defined
$L = \Cov^{-1}$.
%

\subsection{Hamiltonian Dynamics}
The state space is now extended via the introduction of  an auxiliary `velocity' $v\in\Hs$;
the original argument $x\in\Hs$ can be thought of as `location'. We consider the `total energy' function:
\begin{equation}
\label{eq:energy0}
H(x,v;M) =\Phi(x)+\tfrac{1}{2}\,\langle x, Lx \rangle +
\tfrac{1}{2}\,\langle v, M v\rangle \ ,\quad x\in\Hs\ ,
\end{equation}
for a user-specified `mass' operator $M$ such that $M^{-1}$ is a well-defined covariance operator on $\Hs$.
We define the distribution on $\Hs\times \Hs$:
\begin{equation*}
\Pi(x,v) \propto \exp\{ - H(x,v;M) \}  = \exp\{-\Phi(x)-\tfrac{1}{2}\,\langle x, Lx \rangle -
\tfrac{1}{2}\,\langle v, M v\rangle \} \ .
\end{equation*}
Note that under $\Pi(x,v)$ we have $v\sim N(0,M^{-1})$.
%
The Hamiltonian dynamics on $\Hs=\R^{N}$ are defined as follows:
\begin{equation*}
\frac{dx}{dt}=M^{-1}\frac{\partial H}{\partial v}  \ ,\quad
M\,\frac{dv}{dt}=-\frac{\partial H}{\partial x} \ ,
\end{equation*}
or, equivalently:
\begin{equation}
\label{eq:hamilton}
\frac{dx}{dt}=v\ ,\quad  M\,\frac{dv}{dt}=-Lx -\delta \Phi(x) \ .
\end{equation}
Hamiltonian equations preserve the total energy,
and in a probabilistic context they have $\Pi(x,v)$ as their invariant distribution.
The proof (for the case $\Hs=\R^{N}$) under regularity conditions (see e.g.\@~\cite{duna:87})
is straightforward and based on the fact
that the solution operator  of (\ref{eq:hamilton}) is volume-preserving and energy-preserving.
One motivation for the selection of mass matrix $M$
could be that directions of higher variance under the target $\Pi(dx)$ should give higher
marginal velocity variances and vice versa. In particular,
selecting $M=L$ seems optimal in the case when $\Phi\equiv 0$ and the target distribution is simply $N(0,\Cov)$, as we
effectively
transform the target to $N(0,I)$ and equalize all marginal variances.

Caution is needed when $\Hs$ is an infinite-dimensional Hilbert space.
Following \cite{besk:11}, one now is forced to choose the mass matrix
$M = L$ to end up with a well-defined algorithm.
%
In this case the energy function will be:
\begin{equation}
\label{eq:energy}
H(x,v) =\Phi(x)+\tfrac{1}{2}\,\langle x, Lx \rangle +
\tfrac{1}{2}\,\langle v, L v\rangle \ ,\quad x\in\Hs\ ,
\end{equation}
and the Hamiltonian equations will become:
\begin{equation}
\label{eq:hamilton2}
\frac{dx}{dt}=v\ ,\quad \frac{dv}{dt}=-x-\mathcal{C}\,\delta\Phi(x)\ .
\end{equation}
Showing in infinite dimensions that a solution operator of (\ref{eq:hamilton2}) exists and
preserves $\Pi(x,v)$ is harder: an analytical proof is given in \cite{besk:11},
under carefully formulated assumptions on $\Cov$ and $\Phi$.
As the differential equations in (\ref{eq:hamilton2}) cannot be solved analytically,
HMC solves them numerically and then uses an accept/reject step to correct for violating preservation of
total energy (and invariance of $\Pi(x,v)$).

\subsection{Standard HMC on $\Hs=\R^{N}$} The standard HMC algorithm developed in \cite{duna:87} discretises
the Hamiltonian equations (\ref{eq:hamilton}) via a leapfrog scheme (we work under the selection $M=L$):
\begin{align}
v_{h/2} &= v_{0} - \tfrac{h}{2}\,x_0  - \tfrac{h}{2}\,\,\mathcal{C}\,\delta
\Phi(x_0)\ , \nonumber\\
x_h &= x_0 + h\,v_{h/2} \ , \label{eq:lp}\\
v_h &= \,v_{h/2} - \tfrac{h}{2}\,x_h-\tfrac{h}{2}\,\,\mathcal{C}\,\delta\Phi(x_h)
\ , \nonumber
\end{align}
giving rise to an operator $(x_0, v_0)\mapsto (x_h, v_h) = \psi_h(x_0,v_0)$ which is volume-preserving,
and has the symmetricity property $\psih(x_h,-v_h) =(x_0,-v_0)$. HMC looks at Hamiltonian dynamics
up to some time horizon $T>0$, via the synthesis of
\begin{equation}
\label{eq:I}
I=\lfloor\tfrac{T}{h}\rfloor
\end{equation}
leapfrog steps, so we define $\psiT$ to be the synthesis of $I$ mappings $\psih$.
The standard HMC is given in Table \ref{tab:sHMC} ($\mathcal{P}_x$ denotes projection on the $x$-argument).
Due to the  properties of the leapfrog operator mentioned above, it is easy to verify (\cite{duna:87})
that under regulatory conditions the employed acceptance probability provides Markov dynamics with invariant distribution $\Pi(x)$
in (\ref{eq:target}).

%
\begin{table}[!h]
\begin{flushleft}
\medskip
\hrule
\medskip
{\itshape HMC on $\R^N$:}
{\itshape
\vspace{0.2cm}
\begin{enumerate}

\item[(i)] Start with an initial value $x^{(0)}\in \R^N$ and set $k=0$. \vspace{0.2cm}

\item[(ii)] Given $x^{(k)}$ sample  $v^{(k)} \sim N(0,\mathcal{C})$ and propose
$$x^{\star}=\mathcal{P}_x\,\psiT(x^{(k)},v^{(k)})\ .$$
\item[(iii)] Calculate the acceptance probability
\begin{equation}
\label{eq:saccc}
a(x^{(k)},v^{(k)})= 1\wedge \exp\{-\Delta H(x^{(k)},v^{(k)})\}
\end{equation}
for $\Delta H(x,v) = H(\psiT(x,v))-H(x,v)$. \vspace{0.2cm}
\item[(iv)] Set $x^{(k+1)}=x^{\star}$ with probability $a$; otherwise set
$x^{(k+1)}=x^{(k)}.$ \vspace{0.2cm}

\item[(v)] Set $k \to k+1$ and go to (ii).

\end{enumerate}}
\medskip
\hrule
\end{flushleft}
\caption{HMC on $\R^N$, with target  $\Pi(x)$ in (\ref{eq:target}).}
\label{tab:sHMC}
\end{table}

\subsection{Advanced HMC on $\Hs$}

Applying the standard algorithm in Table \ref{tab:sHMC} on an $N$-dimensional projection of the SDE pathspace
would give an algorithm for which: the proposal $x^{\star}$ would become an increasingly inappropriate candidate
for a sample from the target with increasing
$N$ (\cite{besk:11}); thus, the acceptance probability would vanish with increasing $N$, assuming parameters $h$, $T$ were kept fixed.
Indeed, employing the mapping $\psiT$ for elements of the pathspace would project Brownian motion paths to paths of the wrong quadratic
variation which would then necessarily have had acceptance probability $0$ (see \cite{besk:08} for an analytical illustration in the case of MALA);
the results in \cite{besk:13} suggest that one must decrease the step-size $h$ as $\mathcal{O}(N^{-1/4})$ to control the acceptance probability for increasing
$N$.
The advanced HMC algorithm avoids this degeneracy by exploiting the definition  of the target as a change of measure from a
Gaussian law. The development below follows \cite{besk:11}; a similar splitting of the Hamiltonian equations as the one showed
below is also used in \cite{shah:12}, but in a different context.

Hamiltonian equations (\ref{eq:hamilton2}) are split into the following two equations:
\begin{equation}
\label{eq:s1}
\frac{dx}{dt}=0\ ,\quad\frac{dv}{dt}=-\Cov\,\delta \Phi(x)\ ;
\end{equation}
\begin{equation}
\label{eq:s2} \frac{dx}{dt}=v\ ,\quad\frac{dv}{dt}=-x \ .
\end{equation}
Notice that both equations can be solved analytically.
We construct a numerical integrator for (\ref{eq:hamilton2}) by synthesizing steps  on (\ref{eq:s1})
and (\ref{eq:s2}). Analytically, we define the solution operators of (\ref{eq:s1}) and (\ref{eq:s2}):
\begin{equation}
\label{eq:xi1} \Xi_t(x,v) = (x,\,v-t\,\Cov\,\delta\Phi(x))\ ;
\end{equation}
\begin{equation}
\label{eq:xi2}
\tilde{\Xi}_t(x,v) = \big(\cos(t)\,x+\sin(t)\,v,\,
-\sin(t)\,x+\cos(t)\,v\big)\ .
\end{equation}
The numerical integrator for (\ref{eq:hamilton2}) is defined as follows:
\begin{equation}
\label{eq:psih}
\Psi_h = \Xi_{h/2}\circ\XiL\circ\Xi_{h/2}\ ,
\end{equation}
for small $h>0$ and $h^*$ a function of $h$ defined in (\ref{eq:hast}) below.
We can synthesize steps up to some time horizon $T$. Defining $I$ as in (\ref{eq:I}),
$\PsiT$ will correspond to the synthesis of $I$ steps $\Psi_h$.
%
%
$\PsiT$ will provide the proposals  for the MCMC steps.

\begin{rem}
Critically, operators $\Xi_t(x,v)$, $\tilde{\Xi}_t(x,v)$  have the property that they preserve the absolute continuity
properties of an input random pair $(x,v)$ distributed according to the Gaussian law:
\begin{equation}
\label{eq:QQ}
Q_0(x,v) \propto \exp\{-\tfrac{1}{2}\langle x, Lx \rangle - \tfrac{1}{2}\langle v , Lv \rangle  \} \ ,
\end{equation}
(so, also of any other distribution absolutely continuous w.r.t.\@ $Q_0$). This is obvious for $\tilde{\Xi}_t(x,v)$ as it defines
a rotation, so this map is in fact invariant for $Q_0$. Then, as illustrated with Proposition \ref{prop:tran},
Assumption \ref{ass:ass} guarantees precisely
that also $\Xi_{t}(x,v)$ preserves  absolute continuity of $Q_0$. We re-emphasize here, that the solver employed for the
standard HMC algorithm would generate proposals of a distribution singular to the target distribution.
\end{rem}

We will use $h^{\ast}$ such that:
\begin{equation}
\label{eq:hast}
\cos(h^{*}) = \tfrac{1-h^2/4}{1+h^2/4} \ ,
\end{equation}
though any choice is allowed.
For this choice, it can be easily checked that the integrator $(x_0,v_0)\mapsto\Psi_h(x_0,v_0)=:(x_h,v_h)$
can be equivalently expressed as:
\begin{align}
v_{h/2} &= v_{0} - \tfrac{h}{2}\,\frac{x_0+x_h }{2}  - \tfrac{h}{2}\,\,\mathcal{C}\,\delta
\Phi(x_0)\ , \nonumber\\
x_h &= x_0 + h\,v_{h/2} \ , \label{eq:lpf}\\
v_h &= \,v_{h/2} - \tfrac{h}{2}\,\frac{x_0+x_h}{2}-\tfrac{h}{2}\,\,\mathcal{C}\,\delta\Phi(x_h)
\ , \nonumber
\end{align}
which can now be interpreted as a semi-implicit-type integrator of (\ref{eq:hamilton2}).
Under the interpretation (\ref{eq:lpf}), the justification for
the choice (\ref{eq:hast}) is that it delivers an integrator $\Psi_h$
that carries out steps of similar size, $h$, in the $x$ and $v$ directions, in accordance with standard HMC.

The complete algorithm is determined in Table \ref{tab:HMC}.


%
\begin{table}[!h]
\begin{flushleft}
\medskip
\hrule
\medskip
{\itshape HMC on Hilbert space $\Hs$:}
{\itshape
\vspace{0.2cm}
\begin{enumerate}

\item[(i)] Start with an initial value $x^{(0)}\sim \Pi_0\equiv N(0,\Cov)$ and set $k=0$. \vspace{0.2cm}

\item[(ii)] Given $x^{(k)}$ sample  $v^{(k)} \sim N(0,\mathcal{C})$ and propose
$$x^{\star}=\mathcal{P}_x\,\PsiT(x^{(k)},v^{(k)})\ .$$
\item[(iii)] Consider
\begin{equation}
\label{eq:accc}
a(x^{(k)},v^{(k)})= 1\wedge \exp\{-\Delta H(x^{(k)},v^{(k)})\}
\end{equation}
for $\Delta H(x,v) = H(\PsiT(x,v))-H(x,v)$. \vspace{0.2cm}
\item[(iv)] Set $x^{(k+1)}=x^{\star}$ with probability $a$; otherwise set
$x^{(k+1)}=x^{(k)}.$ \vspace{0.2cm}

\item[(v)] Set $k \to k+1$ and go to (ii).

\end{enumerate}}
\medskip
\hrule
\end{flushleft}
\caption{HMC on $\Hs$, with target  $\Pi(x)$ in (\ref{eq:target}).}
\label{tab:HMC}
\end{table}
%
\begin{rem}
The acceptance probability
in the table is at the moment defined only formally, as $H(x,v)=\infty$  a.s.. To see that,
notice that using the Karhunen-Lo\`eve expansion in (\ref{eq:KL-exp}) for $x\sim \Pi_0$ we have
 $\langle x, L x\rangle\equiv \sum_{p=1}^{\infty}\xi_p^2$, for $\xi_p$ iid $N(0,1)$. We re-express the
acceptance probability in the following section in a way that illustrates that the difference
$\Delta H(x,v) = H(\PsiT(x,v))-H(x,v)$
is a.s.\@ well-defined; for the $N$-dimensional projection used in practice one
could still use directly the expression  $\Delta H(x,v) = H(\PsiT(x,v))-H(x,v)$
as each of the two $H$-terms will grow as $\mathcal{O}(N)$.
\end{rem}

\begin{rem}
We will not prove  the existence of a solution for the Hamiltonian equations on Hilbert space (\ref{eq:s1})-(\ref{eq:s2})
or that the solution would preserve $\Pi(x,v)$ as such proofs would require a level and amount of technicalities out of the scope of the paper.
In Section \ref{sec:validity} below we will prove the validity of the algorithm in Table \ref{tab:HMC} which uses directly the numerical integrators of
these equations in (\ref{eq:xi1})-(\ref{eq:xi2}). This seems to suffice from a practical point of view: our proof below indicates that the algorithm
will not collapse as $N\rightarrow\infty$ but will converge to a limit,
with $N$ being the dimension of the vector used instead of complete infinite-dimensional diffusion paths when running the algorithms on a personal computer;
then, for a fixed finite dimension $N$ one can resort to properties of finite-dimensional Hamiltonian equations to justify that, under standard regulatory conditions,
they will indeed preserve the $N$-dimensional target distribution, thus one can attain average acceptance probabilities arbitrarily close to $1$ by decreasing the step-size
$h$.
\end{rem}

\subsection{Proof of Validity for Advanced HMC}
\label{sec:validity}
%
We consider the Gaussian product measure
$Q_0 = N(0,\Cov)\otimes N(0,\Cov)$ on $\Hs\times \Hs$ as in (\ref{eq:QQ}) and the bivariate
distribution $Q$ via the change of measure:
\begin{equation*}
Q(dx,dv) = \exp\{-\Phi(x)\}\,Q_0(dx,dv)\ .
\end{equation*}
We also consider the sequence of probability measures on $\Hs\times \Hs$:
\begin{equation*}
Q^{(i)} = Q\circ \PsiInv{i}\ ,\quad 1\le i\le I\,
\end{equation*}
the sequence $(x_i,v_i) = \Psi_h^{\,i}(x_0,v_0)$, and set:
\begin{equation*}
g(x) := -\Cov^{1/2}\,\delta \Phi(x)\ ,\quad x\in \Hs \ .
\end{equation*}
Note that under Assumption \ref{ass:ass}, $g(x)$ is a well-defined element of the Hilbert space $\Hs$
a.s.\@ under $\Pi_0$. Using Proposition \ref{prop:tran},  we can now prove the following:
%
%
\begin{prop}
\label{prop:rec}
We have that:
\begin{equation*}
\frac{dQ^{(i)}}{dQ_0}(x_i,v_i) = \frac{dQ^{(i-1)}}{dQ_0}(x_{i-1},v_{i-1}) \cdot
G(x_{i},v_{i})\cdot G(x_{i-1},v_{i-1}+\tfrac{h}{2}\,\Cov^{1/2}g(x_{i-1}))\ \ ,
\end{equation*}
where we have defined:
\begin{equation*}
G(x,v) = \exp\big\{ \langle \tfrac{h}{2}\,g(x), \Cov^{-1/2}v \rangle - \tfrac{1}{2}
  |\tfrac{h}{2}\,g(x)|^2  \big\}\ .
\end{equation*}
\end{prop}
\begin{proof}
We will use the chain rule and Proposition \ref{prop:tran}. Recall that
for any two measurable spaces $(E,\mathcal{E})$, $(E{'},\mathcal{E}{'})$,
probability measures $M$, $M_0$ on $(E,\mathcal{E})$ and
1-1 mapping $F : (E,\mathcal{E}) \mapsto (E{'},\mathcal{E}{'})$, we have the following
identity rule for the Radon-Nikodym derivative:
\begin{equation}
\label{eq:general}
 \frac{d\{M_1\circ F^{-1}\}}{d\{M_0\circ F^{-1}\}}(x) =
\frac{dM_1}{dM_0}(F^{-1}(x))\ .
\end{equation}
We now work as follows. Following the definition of $\Psi_h$ from (\ref{eq:psih}), we have the
equality of probability measures:
\begin{equation*}
 Q^{(i)} = Q^{(i-1)}\circ \Xi_{h/2}^{-1}\circ \XiL^{-1} \circ \Xi_{h/2}^{-1}\ .
\end{equation*}
Thus, we have that:
\begin{align}
\frac{dQ^{(i)}}{dQ_0}(x_i,v_i) &= \frac{d\, \{ Q^{(i-1)}\circ \Xi_{h/2}^{-1}\circ \XiL^{-1} \circ \Xi_{h/2}^{-1}\} }{dQ_0}\,(x_i,v_i)  \nonumber \\
&=
\frac{d\, \{ Q^{(i-1)}\circ \Xi_{h/2}^{-1}\circ \XiL^{-1} \circ \Xi_{h/2}^{-1}\} }
{d\,\{ Q_0\circ \Xi_{h/2}^{-1}  \}}\,(x_i,v_i)\times\frac{d\,\{ Q_0\circ \Xi^{-1}_{h/2}\}}{dQ_0}(x_i,v_i)
 \nonumber
\\
 &=
\frac{d\, \{ Q^{(i-1)}\circ \Xi_{h/2}^{-1}\circ \XiL^{-1}\} }
{dQ_0}\,(\Xi_{h/2}^{-1}(x_i,v_i))\times G(x_i, v_i)\ ; \label{eq:tat}
\end{align}
we have used the chain rule in the second line, then (\ref{eq:general}) and Proposition \ref{prop:tran}
(in this case $v_0\equiv \frac{h}{2}\,g(x)$) in the third line.
Using the fact that $Q_0\circ \XiL^{-1}\equiv Q_0$ and that also
$(\XiL^{-1}\circ \Xi_{h/2}^{-1})(x_i,v_i)\equiv \Xi_{h/2}(x_{i-1},v_{i-1})$ we have that:

\begin{align*}
\frac{d\, \{ Q^{(i-1)}\circ \Xi_{h/2}^{-1}\circ \XiL^{-1}\} }
{dQ_0}\,(\Xi_{h/2}^{-1}(x_i,v_i)) \equiv
\frac{d\, \{ Q^{(i-1)}\circ \Xi_{h/2}^{-1} \} }
{dQ_0}\,(\Xi_{h/2}(x_{i-1},v_{i-1})) \ .
\end{align*}
Finally, working as in (\ref{eq:tat}) we have that:
\begin{align*}
\frac{d\, \{ Q^{(i-1)}\circ \Xi_{h/2}^{-1} \} }
{dQ_0}\,&(\Xi_{h/2}(x_{i-1},v_{i-1})) = \\ &=
\frac{d Q^{(i-1)} }
{dQ_0}\,(x_{i-1},v_{i-1}) \times
\frac{d\,\{ Q_0\circ \Xi^{-1}_{h/2}\}}{dQ_0}(\Xi_{h/2}(x_{i-1},v_{i-1})) \\
&= \frac{d Q^{(i-1)} }
{dQ_0}\,(x_{i-1},v_{i-1}) \times
G(\Xi_{h/2}(x_{i-1},v_{i-1}))\ .
\end{align*}
The definition of $\Xi_{h/2}$ gives $G(\Xi_{h/2}(x_{i-1},v_{i-1}))\equiv G(x_{i-1},v_{i-1}+\tfrac{h}{2}\,\Cov^{1/2}g(x_{i-1}))$. Following the calculation
from (\ref{eq:tat}) we have now proven the requested result.

%
\end{proof}

Thus, using Proposition \ref{prop:rec} iteratively we have now obtained that:
\begin{equation}
\label{eq:exp1}
\frac{dQ^{(I)}}{dQ_0}(x_I,v_I) = \frac{dQ}{dQ_0}(x_{0},v_{0}) \times \prod_{i=1}^{I}
G(x_{i},v_{i})\, G(x_{i-1},v_{i-1}+\tfrac{h}{2}\,\Cov^{1/2}g(x_{i-1}))\ \ .
\end{equation}
Now, following the definition of $\Psi_{h}$ in (\ref{eq:psih}), we set:
\begin{align*}
v^{-}_{i-1} &= \mathcal{P}_v\,\Xi_{h/2}(x_{i-1},v_{i-1}) \equiv v_{i-1} + \tfrac{h}{2}\,\Cov^{1/2}g(x_{i-1})\ ; \\
v^{+}_{i} &= \mathcal{P}_v\,(\,\XiL\circ \Xi_{h/2}(x_{i-1},v_{i-1}) \,) \equiv v_i - \tfrac{h}{2}\,\Cov^{1/2}g(x_{i})\ .
\end{align*}
($\mathcal{P}_v$ denotes projection onto the $v$-argument.)
Using these definitions, for any $h, h^{\ast}>0$ we have that:
\begin{align*}
\log\{ &\,G(x_{i},v_{i})\, G(x_{i-1},v_{i-1}+\tfrac{h}{2}\,\Cov^{1/2}g(x_{i-1}))\,\} = \\
&=  \langle \tfrac{h}{2}\,g(x_i), \Cov^{-1/2}v_i \rangle - \tfrac{1}{2}|\tfrac{h}{2}g(x_i)|^{2}
+\langle \tfrac{h}{2}\,g(x_{i-1}), \Cov^{-1/2}v_{i-1} \rangle +  \tfrac{1}{2}|\tfrac{h}{2}g(x_{i-1})|^{2}
\\
&=\tfrac{1}{2}\,\langle v_i, Lv_i \rangle - \tfrac{1}{2}\,\langle v_i^{+}, Lv_i^{+} \rangle
- \tfrac{1}{2}\,\langle v_{i-1}, Lv_{i-1} \rangle + \tfrac{1}{2}\,\langle v_{i-1}^{-}, Lv_{i-1}^{-} \rangle
 \\
& =
\tfrac{1}{2}\,\langle x_{i}, Lx_{i} \rangle + \tfrac{1}{2}\,\langle v_{i}, L v_{i}\rangle
- \tfrac{1}{2}\,\langle x_{i-1}, Lx_{i-1} \rangle - \tfrac{1}{2}\,\langle v_{i-1}, L v_{i-1}\rangle\ .
\end{align*}
The last equation is due to the mapping $(x_{i-1}, v_{i-1}^{-})\mapsto (x_i,v_i^{+})$ corresponding
to the modulus-preserving rotation $\XiL$.
Thus, we can rewrite (\ref{eq:exp1}) as follows:
\begin{equation}
\label{eq:exp2}
\frac{dQ^{(I)}}{dQ_0}(x_I,v_I) = \exp\{\Delta H(x_0,v_0) - \Phi(x_I)\}\ .
\end{equation}
The above expression will be used at proving the main result below.
\begin{rem}
\label{rem:prop}
The operator $\Psi_h$ (thus, also $\PsiT$)  has the following properties:
\begin{itemize}
 \item[i)] $\Psi_h$ is symmetric, that is $ \Psi_h \circ S\circ \Psi_h = S $ where $S(x,v) = (x,-v)$.
\item[ii)]  $\Psi_h$ is (formally) volume-preserving, as it preserves volume when $\Hs\equiv \R^{d}$.
\end{itemize}
\end{rem}
\begin{thm}
The Markov chain with transition dynamics specified in Table~\ref{tab:HMC}
has invariant distribution $\Pi(x)$ in (\ref{eq:target}).
\end{thm}
\begin{proof}
Assuming stationarity, so $(x_0,v_0)\sim Q$, we can write for the next position, $x'$, of the Markov chain (recall that $(x_I,v_I)=\PsiT(x_0,v_0)$):
\begin{equation*}
x'  = \mathrm{I}\,[\,U\le a(\PsiInv{I}(x_I,v_I))\,]\,x_{I} + \mathrm{I}\,[\,U>a(x_0,v_0)\,]\,x_0\ ,
\end{equation*}
for a uniform random variable $U\sim \mathrm{Un}\,[0,1]$. Let $f:\Hs\mapsto \R$ be bounded and continuous.
We need to prove that:
\begin{equation*}
\Exp\,[\,f(x')\,] = \Exp\,[\,f(x_0)\,]\ .
\end{equation*}
Integrating out $U$ from above we get:
\begin{equation}
\label{eq:final}
\Exp\,[\,f(x')\,]  = \Exp[\,f(x_I)\,a(x_0,v_0)\,] - \Exp[\,f(x_0)\,a(x_0,v_0)\,] +
\Exp\,[\,f(x_0)\,]\ .
\end{equation}
Note now that (here, we need to stress the integrators in expectations/integrals and will
 show them explicitly as a subscript of $\Exp$):
\begin{align}
\Exp[\,f(x_I)\,a(x_0,v_0)\,]& = \Exp_{\,Q^{(I)}}[\,f(x_I)\,a(\PsiInv{I}(x_I,v_I))\,]
\nonumber\\
 &\!\!\stackrel{(\ref{eq:exp2})}{=}  \Exp_{\,Q_0}[\,f(x_I)\,a(\PsiInv{I}(x_I,v_I))\,
e^{\Delta H(\PsiInv{I}(x_I,v_I))-\Phi(x_I)}\,] \nonumber \\
&=  \Exp_{\,Q_0}[\,f(x_I)\,(\,1\wedge e^{\Delta H(\PsiInv{I}(x_I,v_I))}\,)\,e^{-\Phi(x_I)}\,]\nonumber \\
&= \Exp_{\,Q}[\,f(x_I)\cdot 1\wedge e^{\Delta H(\PsiInv{I}(x_I,v_I))}\,]\ \nonumber \\
&= \Exp_{\,Q}[\,f(x_I)\cdot 1\wedge e^{\Delta H(\PsiInv{I}(x_I,-v_I))}\,] \ .\label{eq:aa}
\end{align}
(For the last equation, notice that $(x_I,v_I)$ and $(x_I,-v_I)$ have the same law $Q$.)
Now, that due to the symmetricity property $\PsiT \circ S\circ \PsiT = S $ of the leapfrog operator
in Remark \ref{rem:prop} we have that $ \PsiInv{I}\circ S = S\circ \PsiT $. Thus, we have:
\begin{align*}
\Delta &H(\PsiInv{I}(x_I,-v_I))) = \Delta H(S\circ \PsiT(x_I,v_I))) \\
&=  H(S(x_I,v_I)) - H(S\circ \PsiT(x_I,v_I))
 \equiv -\Delta H(x_I,v_I)\ ,
\end{align*}
where is the last equation we used the fact that $H\circ S=H$ due to the energy $H$ being quadratic in
the velocity $v$.
Thus, using this in (\ref{eq:aa}), we have that:
\begin{equation}
\label{eq:fff}
\Exp[\,f(x_I)\,a(x_0,v_0)\,] = \Exp_{\,Q}[\,f(x_I) a(x_I,v_I)\,] \equiv \Exp[\,f(x_0)\,a(x_0,v_0)\,] \ .
\end{equation}
So, from (\ref{eq:final}), the proof is now complete.
\end{proof}
\begin{rem}
The demonstration of validity of standard HMC \cite{duna:87} does not require the recursive calculation of the
forward density (\ref{eq:exp2}) as it exploits the preservation of volume (unit Jacobian) for the mapping
$(x_0, v_0)\mapsto \psiT(x_0,v_0)$ to directly prove the analogue to~(\ref{eq:fff}).  So, finding (\ref{eq:exp2})
overcomes the difficulty of making sense of a Jacobian for the transform $\PsiT$ on the infinite-dimensional Hilbert space.
\end{rem}

%
\subsection{Advanced MCMC Samplers on Pathspace}

A number of MCMC algorithms corresponding to an upgrade of
standard RWM, MALA and HMC on the infinite-dimensional pathspace are now available.
So far in Section \ref{sec:advHMC} we have defined HMC on pathspace under Assumption \ref{ass:ass}. In \cite{besk:08}  the definition of advanced MALA on the pathspace is provided;
also advanced RWM on pathspace was defined in that paper. We now briefly review these two local-move algorithms.

The starting point for MALA is a Langevin SDE with drift $\frac{1}{2}\,\Cov\,\delta \log \Pi(x)$ and
coefficient $\Cov^{1/2}$, that is, after a calculation on the drift:
%
%
%
\begin{equation}
\label{eq:Langevin}
\frac{dx}{dt} = -\tfrac{1}{2}\,x - \tfrac{1}{2}\,\Cov\,\delta\Phi(x) + \Cov^{1/2}\,\frac{dw}{dt} \ .
\end{equation}
In an Euclidean setting $\{w_t\}$ denotes a standard Brownian motion, whereas in the pathspace
it denotes a cylindrical Brownian motion. In both cases, the process can be easily understood via the distribution of  it's increments, as
$\Cov^{1/2}\tfrac{(w_{t+s}-w_t)}{\sqrt{s}}\sim  N(0,\Cov)$.
On pathspace, the SDE (\ref{eq:Langevin}) is
shown in \cite{besk:08} to have invariant distribution $\Pi$ under Lipschitz continuity and
absolute boundedness assumptions on $\delta \Phi$. In the practically interesting case of nonlinearity,
this SDE cannot be solved analytically. So, a proposal can be derived
via the following Euler-type scheme on (\ref{eq:Langevin}) for an finite increment $\Delta t >0$:
\begin{equation}
\label{eq:MALAprop}
x^{\ast} - x = -{\Delta t}\,(\theta\,\tfrac{x^{\ast}}{2}+(1-\theta)\tfrac{x}{2})
- \tfrac{\Delta t}{2}\,\Cov\,\delta\,\Phi(x) + \sqrt{\Delta t}\,N(0,\Cov)\ .
\end{equation}
Standard MALA is derived from an explicit Euler scheme with $\theta=0$ and advanced pathspace MALA
from a semi-implicit scheme with $\theta=1/2$.
Contrasting (\ref{eq:MALAprop}) with the leapfrog steps in (\ref{eq:lp}) and (\ref{eq:lpf}), one can easily check
that standard (resp.\@ advanced) MALA is a particular case of standard (resp.\@ advanced) HMC when choosing $h = \sqrt{\Delta t}$ and a single leapfrog step $I=1$.

%
%

Finally, a RWM algorithm on pathspace is derived in \cite{besk:08} via proposal (\ref{eq:MALAprop})
for $\theta=1/2$ but also with omitting the nonlinear term $\Cov\,\delta \Phi(x)$. That is, the proposal for advanced RWM is:
\begin{equation*}
x^{\ast} = \rho\,x + \sqrt{1-\rho^2}\,N(0,\Cov)\ ,
\end{equation*}
with parameter $$\rho = \frac{1-\frac{\Delta t}{4}}{1+ \frac{\Delta t}{4}}\ .$$ The Metropolis-Hastings
acceptance probability for this proposal
(see \cite{besk:08}) is reminiscent of the one for standard RWM, namely $1\wedge \tfrac{\Pi(x^{\ast})}{\Pi(x)}$, which also explains the interpretation
of this algorithm as `advanced RWM'.
Table \ref{tab:MCMC} summarises the three pathspace samplers looked at in this paper together with their standard versions
for finite-dimensional spaces.
\setlength\extrarowheight{4pt}
\begin{table}[h!]
\begin{tabular}{|l|l|l|}
\hline
Algorithm & Pathspace Proposal & Standard Proposal   \\
\hline
HMC       &  $x^{\ast} = \mathcal{P}_x \PsiT(x,v)$       &   $x^{\ast} = \mathcal{P}_x\psiT(x,v)$      \\
MALA      &     $x^{\ast} = \rho\,x + \sqrt{1-\rho^2}\,v - \tfrac{\Delta t}{2}\,\Cov\,\delta \Phi(x)$   &
$ x^{\ast} = (1-\tfrac{\Delta t}{2})x - \frac{\Delta t}{2}\,\Cov\,\delta\Phi(x) + \sqrt{\Delta t}\,v$    \\
RWM       &   $x^{\ast} = \rho\,x + \sqrt{1-\rho^2}\,v$   &
$x^{\ast} = x + \sqrt{\Delta t}\,v$    \\
\hline
\end{tabular}
\caption{MCMC algorithms on pathspace together with their standard versions. In all cases $v\sim N(0,\Cov)$.
HMC for $I=1$ and $h=\sqrt{\Delta t}$ coincides with MALA.}
\label{tab:MCMC}
\end{table}
\section{HMC Superiority in an Analytical Study}
\label{sec:OUanalytical}
In the applications that we show in the next sections HMC appears to be much more efficient than MALA,
even if they both use the same information about the target distribution in the form of
 the derivative $\delta \log \Pi$. So, the synthesis
of deterministic steps for HMC, by avoiding random-walk-type behaviour for the Markov chain dynamics,
seems to be providing significant computational advantages for HMC.
We will illustrate this analytically in this section for the case of a linear
target distribution corresponding to an Ornstein-Uhlenbeck (OU) diffusion process. In particular, we will
show that HMC gains orders of complexity compared to MALA and RWM
when MALA itself does not gain a complexity benefit over RWM.
So, an important message derived here is that designated use of information on the derivative
can have significant effect on the efficiency
of such MCMC methods.

We will consider the OU bridge:
\begin{align}
dX_u &= -\kappa X_u\, dt + dB_u \ , \nonumber\\
\quad &X_0=X_{\ell} = 0   \ ,\label{eq:OU}
\end{align}
for reversion parameter $\kappa>0$ and path-length $\ell>0$.
From Girsanov's theorem (\cite{okse:03}) we get that the target distribution
is defined on the Hilbert space $\Hs = L^{2}([0,\ell],\R)$ and is expressed as in the general form (\ref{eq:target}),
so that:
\begin{equation}
\label{eq:OUtarget}
\frac{d\Pi}{d\Pi_0}(x)= \exp\{-\Phi(x)\}\ ; \quad \Pi_0 =N(0,\Cov^{bb})\,,\,\,
\Phi(x)= \tfrac{\kappa^2}{2}\int_{0}^{\ell}x^2(u)du+c\ ,
\end{equation}
for some constant $c\in\R$,
with $N(0,\Cov^{bb})$ the distribution of a Brownian bridge with $x(0)=x({\ell})=0$.
We will look at the complexity of pathspace samplers
as a function of the length $\ell$ of the bridge.
Our main result summarises the mixing times as follows:
\begin{align}
\textrm{RWM}&:\,\,\, \mathcal{O}(\ell^{2})\ ;\nonumber\\
\textrm{MALA}&:\,\,\, \mathcal{O}(\ell^{2})\ ;\label{eq:mixx}\\
\textrm{HMC}&:\,\,\,\mathcal{O}(\ell)\ .\nonumber
\end{align}
The notion of mixing time is used here in an informal, practical manner and should not be confused with
analytical definitions of various different
versions of mixing times appearing in the Markov chain literature. In particular, the results below
provide appropriate scalings of the step-sizes for the relevant MCMC samplers as a function of $\ell$
that deliver non-vanishing acceptance probabilities as $\ell$ grows. Then, informal arguments will be used to connect mixing
times with the inverse of such step-sizes.
%

%
\subsection{Acceptance Probability for RWM, MALA, HMC}
The proof in Proposition~\ref{prop:1} below follows closely the derivations in the PhD thesis~\cite{whit:09},
where a similar scaling problem has been considered over the reversion parameter $\kappa$.
We have decided to include an analytical proof in the Appendix for reasons of completeness, but also because we have made
some modifications that make it easier for the reader to follow the derivations.
\subsubsection*{Karhunen-Lo\`eve expansion for Brownian Bridge and OU Bridge}
The Karhunen-Lo\'eve expansion (see Section \ref{sec:gaussian}) of the Gaussian distributions corresponding to
the target OU bridge and the reference Brownian bridge will be used in this section.
In particular, we will use the orthonormal basis $\{\phi_p\}_{p=1}^{\infty}$ of $\Hs$ corresponding to
the eigenfunctions of $\Cov^{bb}$ and make the standard correspondence $x\mapsto \{x_p\}_{p=1}^{\infty}$ between an element $x\in\Hs$
and it's squared summable co-ordinates $x_p=\langle x,\phi_p \rangle $ w.r.t.\@ the  basis $\{\phi_p\}$.
In particular, the eigen-structure $\{\lambda_p, \phi_p\}_{p=1}^{\infty}$ of $\Cov^{bb}$ is specified as follows (see e.g.\@ \cite{whit:09}):
\begin{equation}
\label{eq:BBB}
\lambda_p = \frac{\ell^2}{\pi^2 p^2}\ ; \quad \phi_p(u) = \sqrt{\frac{2}{\ell}}\,\sin(\frac{\pi p u}{\ell})\ .
\end{equation}
Then, the Karhunen-Lo\`eve expansion of the two Gaussian distributions w.r.t.\@ the above basis of sinusoidals is as below (see e.g.\@ \cite{whit:09}):

\begin{equation}
\label{eq:KL}
\textrm{BB:}\,\,\,\,
 x = \sum_{p=1}^{\infty} \frac{\ell}{\pi p}\,\xi_p\,\phi_p\ ; \quad
\textrm{OU Bridge:}\,\,\,\,
x = \sum_{p=1}^{\infty} \frac{1}{\sqrt{ \frac{\pi^2p^2}{\ell^2}+\kappa^2 }}\,\xi_p\,\phi_p\ ,
\end{equation}
where $\{\xi_p\}_{p=1}^{\infty}$ are iid variables from $N(0,1)$.
%

\begin{prop}
\label{prop:1}
Consider the advanced MALA and RWM algorithms described in Table \ref{tab:MCMC}
with target distribution $\Pi$ in (\ref{eq:OUtarget}). If $a=a(x,v)$ is the acceptance
probability of the proposal for current position $x$ and $v\sim N(0,\Cov^{bb})$, then in stationarity ($x\sim \Pi$) we have the following:
\begin{itemize}
 \item If $\Delta t = c / \ell^{2}$ for some constant $c>0$,  then $\lim\sup_{\ell}\Exp\,[\,a\,] > 0$.
\item If $\Delta t = c / \ell^{\epsilon}$ for $\epsilon\in(0,2)$ and a constant $c>0$, then $\lim_{\ell\rightarrow\infty}\Exp\,[\,a\,]= 0$.
\end{itemize}
\end{prop}
\begin{proof}
See the Appendix.
\end{proof}	
We will now derive a corresponding result for HMC. Recall that the step-size for HMC is denoted by $h$
instead of $\Delta t$.
We will show that the scaling $h=c/\ell$ for some constant $c>0$ will control the average acceptance probability.
We work as follows (under the choice $h=c/\ell$).
Each leapfrog step $\Psi_h$, defined via (\ref{eq:psih}), (\ref{eq:hast}), can in this case be written as a linear operator:
\begin{equation*}
\Psi_h  =
\left(
\begin{array}{cc}
 \rho - (1-\rho)\kappa^2\Cov^{bb} & \sqrt{1-\rho^2}\vspace{0.2cm} \\
-\frac{1 - \big(\rho-(1-\rho)\kappa^2\Cov^{bb}\big)^2}{\sqrt{1-\rho^2}} &
\rho - (1-\rho)\kappa^2\Cov^{bb}
\end{array}
\right)
\end{equation*}
where we have set here:
\begin{equation*}
\rho = \frac{1-\frac{h^2}{4}}{1+\frac{h^2}{4}} \ .
\end{equation*}
So, the mapping $\Psi_h$ above, can be equivalently expressed in terms of it's effect
to the $p$th co-ordinates $x_p$, $v_p$ of $x$, $v$ respectively w.r.t.\@ the orthonormal basis $\{\phi_p\}$ in (\ref{eq:BBB})  as follows:
\begin{equation*}
\Psi_{h,p}  =
\left(
\begin{array}{cc}
 \rho - (1-\rho)\kappa^2\lambda_p & \sqrt{1-\rho^2}\vspace{0.2cm} \\
-\frac{1 - \big(\rho-(1-\rho)\kappa^2\lambda_p\big)^2}{\sqrt{1-\rho^2}} &
\rho - (1-\rho)\kappa^2\lambda_p
\end{array}
\right)\ .
\end{equation*}
Powers of the above matrix will be determined via it's eigenstructure.
We will only consider the case when there will be complex eigenvalues, i.e.\@ when:
\begin{equation*}
|\rho-(1-\rho)\,\kappa^2\lambda_p| < 1\ ,
\end{equation*}
as in the alternative scenario  there will be an eigenvalue of modulus greater than
one (since the Jacobian of the above matrix is unit) whose powers will explode rendering the algorithm unstable.
The above is equivalent to requiring that $(4-\frac{c^2}{\ell^2}-
\frac{2\,c^2\,\kappa^2}{p^2\,\pi^2})/(4+\frac{c^2}{\ell^2})$ lies in $(-1,1)$, which can be easily seen
to be guaranteed, for any $\ell\ge \ell_0>0$ and for all $p\ge 1$, under the condition:
\begin{equation}
\label{eq:condition}
 c\,\kappa < 2\pi \ .
\end{equation}
This condition specifies the region of stability (see e.g.\@ \cite{leim:04}) for the discretisation scheme of the Hamiltonian dynamics in our context.
Under (\ref{eq:condition}), we can conveniently write:
\begin{equation}
\label{eq:power}
 \Psi_{h,p}^{I} =
\left(
\begin{array}{cc}
 \cos(\theta_p) & \!\!\! a_p\,\sin(\theta_p)\\
-\tfrac{1}{a_p}\,\sin(\theta_p) &  \!\!\! \cos(\theta_p)
\end{array}
\right)^{I}
\equiv
\left(
\begin{array}{cc}
 \cos(I\theta_p) &  \!\!\! a_p\,\sin(I\theta_p)\\
-\tfrac{1}{a_p}\,\sin(I\theta_p) & \!\!\! \cos(I\theta_p)
\end{array}
\right)
\end{equation}
where we have set:
\begin{equation}
\label{eq:dddd}
\cos(\theta_p)=  \rho-(1-\rho)\,\kappa^2\lambda_p \ ;\quad \sin(\theta_p) = \sqrt{1-\cos^{2}(\theta_p)}\ ;\quad
 a_p = \tfrac{\sqrt{1-\rho^2}}{\sin(\theta_p)} \ .
\end{equation}
We can now derive the following result.
\begin{prop}
\label{prop:2}
Consider the  advanced HMC algorithm in Table \ref{tab:HMC}  with target distribution $\Pi$ in (\ref{eq:OUtarget}).
If $a=a(x,v)$ is the acceptance
probability of the proposal for current position $x$ and $v\sim N(0,\Cov^{bb})$, then in stationarity ($x\sim \Pi$) we have the following:
\begin{itemize}
 \item If $h = c/\ell$ with $c\,\kappa < 2\pi$ then $\lim\inf_{\ell}\Exp\,[\,a\,] > 0$.
\end{itemize}
\end{prop}
\begin{proof}
See the Appendix.
\end{proof}
\noindent So, under the selection $h=c/\ell$, the average acceptance probability is lower bounded by a constant for arbitrarily long bridges.

\begin{rem}
We could now informally connect the above step-sizes that control the average acceptance probability for the advanced RWM, MALA and HMC algorithms with
their mixing times, which will involve their inverses, as stated in (\ref{eq:mixx}). One could think of the effect
of the proposal of each algorithm for increasing $\ell$ on a fixed time-window for a path, say on $[0,\ell_0]$ for some $\ell_0>0$.
For HMC, the synthesis of $I=\lfloor \frac{T}{h} \rfloor$ leapfrog steps will give a proposal moving the whole sub-path on $[0,\ell_0]$ an
$\mathcal{O}(1)$-distance within it's state space. To show that, we ignore for a moment the effect of the nonlinear map $\Xi_{h/2}$ at the
 the leapfrog update in (\ref{eq:psih}) and focus on the synthesis of $I$ linear maps~$\XiL$. This gives:
\begin{equation*}
\XiL^{I} = \left( \begin{array}{cc}
            \cos(Ih^{\ast}) & \sin(Ih^\ast) \\
  	    -\sin(Ih^{\ast}) & \cos(Ih^{\ast})
           \end{array} \right) \longrightarrow
 \left( \begin{array}{cc}
            \cos(T) & \sin(T) \\
  	    -\sin(T) & \cos(T)
           \end{array} \right) \ ,\quad \textrm{as}\,\,\ell\rightarrow \infty \ .
\end{equation*}
The effect of the nonlinear operator $\Xi_{h/2}$ does not have a similarly simple interpretation, but
should not offset the main effect of proposals making $\mathcal{O}(1)$-steps from a current position, for arbitrarily large $\ell$.
Thus, as a function of $\ell$, the mixing time for advanced HMC only corresponds to the order of the number of leapfrog steps, $\mathcal{O}(\ell)$.
For advanced RWM, shown in Table \ref{tab:MCMC}, for $\Delta t = c/\ell^2$ we can express the proposal as:
\begin{equation}
\label{eq:rr}
x^{\ast} = (1 + \mathcal{O}(\ell^{-2}))\,x + \tfrac{\sqrt{c}}{\ell}\,(1+ \mathcal{O}(\ell^{-2}))\,\xi \ .
\end{equation}
Here, due to the random-walk nature of the proposal, the algorithm will have to synthesize $\mathcal{O}(\ell^2)$-steps to
move $\mathcal{O}(1)$-distance from a current position for a fixed point of the sub-path in $(0,\ell_0]$,
thus the $\mathcal{O}(\ell^{2})$-mixing time. Finally, for MALA, one has to refer
to the interpretation of the algorithm as a discretisation of an SDE on the pathspace, expressed in (\ref{eq:Langevin}):
without being rigorous here, advanced MALA essentially carries out steps of size $\Delta t = \mathcal{O}(\ell^{-2})$ along the continuous-time dynamics, thus
will require $1/\Delta t  =\mathcal{O}(\ell^2)$ steps to propagate a point of the sub-path on $[0,\ell_0]$ an $\mathcal{O}(1)$-distance from it's current position.

Of course, a rigorous analysis of mixing times would involve characterising the eigenvalues of the Markov chains, but this is beyond the scope  of this paper.
\end{rem}

\section{Advanced MCMC for Diffusion-Driven Models}
\label{sec:ReducibleDiff}
In this section we return to the general framework of Section \ref{sec:introdif} that can accommodate a wide range of applications. Recall that the statistical model here is driven by the solution $V$ of the SDE (\ref{eq:diffusion})
involving some unknown parameter $\theta$; then, conditionally on $V$ and $\theta$ there are some observations $Y$ via the Lebesgue
density $p(y|v,\theta)$. We will now apply the previously defined advanced MCMC samplers in the context of such diffusion-driven models
to efficiently sample from the path-valued distribution $p(v|y,\theta)$.
Such models provide a generic framework where target distributions
defined as a change of measure from Gaussian laws as in (\ref{eq:target})
will naturally rise due to the probabilistic dynamics
being driven by Brownian motion.
We will first give a brief description of inferential issues
related with such models, before we show details over the applicability
of the advanced MCMC methods in such a context.

The method in \cite{rob:str01} succeeds in breaking down the singularity
between $V$ and the parameters in the diffusion coefficient $\sigma(\cdot;\theta)$ through appropriate use of the Lamperti transform;
see \cite{kal07} and \cite{kal:del:rob11} for extensions. This framework covers the case of
\emph{reducible} diffusions, i.e.\@ SDEs that can be transformed via an 1-1 map to ones of unit diffusion
  coefficient. Now,  \cite{kal:rob:del10} offers some extensions
to irreducible diffusions
utilizing time change transformations; nevertheless the most general framework for irreducible diffusions is offered from
 \cite{ch:pit:she06, goli:08} requiring only invertibility of the diffusion coefficient.
For ease of exposition we will first consider the case of univariate ($d=1$) reducible diffusions  and work
with the reparametrisation recipe of \cite{rob:str01}.
However, advanced HMC algorithms can also be defined under (\ref{eq:coef}) with the
approach of \cite{goli:08,ch:pit:she06}; details are postponed until Section~\ref{sec:GeneralDiff}.

For reducible diffusions we can apply the Lamperti transform on the original SDE model (\ref{eq:diffusion}), i.e.
$V_u\rightarrow \eta(V_u;\theta)=:X_u$ where
$$
\eta(v;\theta)=\int_{V_0}^v\frac{1}{\sigma(z,\theta)}dz
$$
is the antiderivative of $\sigma^{-1}(\cdot;\theta)$.
Depending on the context, $V_0$ can be treated as fixed or as an additional unknown parameter. Assuming that $\sigma(\cdot;\theta)$ is continuously
differentiable, an application of Ito's lemma provides the SDE of the transformed diffusion $X$ as:
\begin{equation}
  \label{eq:diffusionunitvol}
    dX_u=\nu(X_u;\theta)du+dB_u\  ,\quad  X_0=0\ ,
\end{equation}
where
\begin{equation*}
\nu(x;\theta)=\frac{\mu\big(\eta^{-1}(x;\theta);\theta\big)}{\sigma\big(\eta^{-1}(x;\theta);\theta\big)}- \tfrac{1}{2} \sigma^{\prime}\big(\eta^{-1}(x;\theta);\theta\big)\ .
\end{equation*}
%
Girsanov's theorem  gives that (for $\Pi_0$ being the standard Brownian motion on $[0,\ell]$):
\begin{equation*}
\frac{d\Pi}{d\Pi_0}(x|\theta) =
\exp\Big\{\int_{0}^{\ell}\nu\big(x(u);\theta\big)dx(u) -
\tfrac{1}{2}\int_{0}^{\ell}|\nu\big(x(u);\theta\big)|^2\,du\Big\} \ .
\end{equation*}
Thus, for the distribution of  $X$ conditionally on the data $Y$ we have that:
\begin{align}
\label{eq:targetGenDiff}
\frac{d\Pi}{d\Pi_0}(x|y,\theta) &\propto p(y|\eta^{-1}(x;\theta),\theta)\,\frac{d\Pi}{d\Pi_0}(x|\theta) \nonumber\\
 &\propto \exp\{-\Phi(x;y,\theta)\} \ ,
\end{align}
where we have defined:
\begin{align}
\label{eq:Phi}
\Phi(x;y,\theta)= -\int_{0}^{\ell}\nu(x(u);\theta)dx(u)+
\tfrac{1}{2}&\int_{0}^{\ell} \nu^2(x(u);\theta)du\nonumber  \\ &- \log p(y|\eta^{-1}(x;\theta),\theta)\ .
\end{align}

Next, assuming a generic structure for $p(y|v,\theta)$ relevant for practical applications, we will verify Assumption~\ref{ass:ass} and therefore demonstrate that our advanced HMC method is well-defined for such models.

\subsection{Calculation of $\Cov\,\delta \Phi(x) $}

Motivated by the expression in (\ref{eq:Phi}) for $\Phi(x|y,\theta)$
and the structure of the data density $p(Y|V,\theta)$ arising in applications,  we will carry out calculations
assuming the following general form:
\begin{equation}
\Phi(x)  = \alpha(x(u_1),x(u_2),\ldots, x(u_M))+ \beta( I_1, I_2,\ldots, I_L) + \gamma(S_1, S_2,\ldots, S_J)
\label{eq:structure}
\end{equation}
where we have set:
\begin{align*}
I_{l} = \int_{0}^{\ell} z_l(u, x(u))du \,\,\, ,\,1\le l\le L\ ;\quad S_j =
\int_{0}^{\ell}r_j(u,x(u))dx(u) \,\,\, , \,1\le j \le J\  ,
\end{align*}
for positive integers $M, L, J$, times $u_1<u_2<\cdots<u_M$ in $[0,\ell]$
that could be determined by the data $Y$ and functions $\alpha,\beta,\gamma, z_l, r_j$
determined via the particular model. Explicit instances of this structure will be provided
in the example applications in the  sequel.
Here, the target posterior distribution $\Pi(dx)$ is defined on the Hilbert space of squared integrable paths
$\Hs = L^{2}([0,\ell],\R)$ (with appropriate boundary conditions).
The centered Gaussian reference measure $\Pi_0$ will correspond to a Brownian motion (thus, boundary condition $x(0)=0$) or a Brownian Bridge ($x(0)=x(\ell)=0$).
Recall here the specification of the covariance operators $\Cov^{bm}$, $\Cov^{bb}$ and  Cameron-Martin spaces $\Hs_0^{bm}$, $\Hs_0^{bb}$ of a Brownian motion and Brownian bridge respectively in Section \ref{sec:gaussian}. We  make the following definitions, for the relevant range of subscripts:
\begin{gather*}
\alpha_m = \frac{\partial\alpha}{\partial x(u_m)}(x(u_1),x(u_2),\ldots,x(u_{M}))\ ;\quad
\beta_l = \frac{\partial\beta}{\partial I_l}(I_1,I_2,\ldots,I_L)\ ;\\
 \gamma_j = \frac{\partial\gamma}{\partial S_j}(S_1,S_2,\ldots,S_J)\ ;\quad   z'_l = \frac{\partial z_l}{\partial x}\ ; \quad r'_j =  \frac{\partial r_j}{\partial x}\ .
\end{gather*}

\begin{rem}
With a somewhat abuse of notation,
path-elements $\{\Cov^{bm}\delta\Phi(x)\}$, $\{\Cov^{bb}\delta\Phi(x)\}$  found in Proposition \ref{prop:3} below
are obtained (at least for the terms in $\Phi(x)$ involving stochastic integrals)
by recognising that the finite-difference $N$-dimensional algorithm used in practice
corresponds to applying the finite-difference
scheme on the Hilbert-space-valued algorithm employing precisely the shown paths
$\{\Cov^{bm}\delta\Phi(x)\}$, $\{\Cov^{bb}\delta\Phi(x)\}$ within it's specification.
(So, here $\delta \Phi(x)$ corresponds to a
variational derivative only formally.) This remark applies also to a similar result shown in
Proposition \ref{prop:transform} in a subsequent section.
\end{rem}

\begin{prop}
\label{prop:3}
For the functional $\Phi(x)$ given in (\ref{eq:structure}), for any
$x\in\Hs$:
\begin{align*}
\big(&\Cov^{bm}\delta\Phi(x)\big)(u) = \sum_{m=1}^{M}\alpha_m\cdot
\big(\,u\,\mathbb{I}\,[\,u < u_m\,] + u_m\,\mathbb{I}\,[\,u\ge u_m\,]\,\big) \\
&+ \sum_{l=1}^{L}\beta_l \cdot \Big( u\int_{0}^{\ell}z'_l(v,x(v))dv - \int_{0}^{u}\int_{0}^{s}z'_l(v,x(v))dv\,ds \Big) \\
&+ \sum_{j=1}^{J}\gamma_j \cdot \Big( u\,\big(\, r_j(\ell,x(\ell))+\int_{0}^{\ell}dq_j(v)\,\big) - \int_{0}^{u}\int_{0}^{s}dq_j(v)\,ds\Big)
\ , \quad u\in[0,\ell]\ ,
\end{align*}
for the integrator  $$dq_j(v) = r'_j(v,x(v))dx(v) - dr_j(v,x(v))\ .$$
Also:
\begin{align*}
\big(\Cov^{bb}\delta\Phi(x)&\big)(u)=
\sum_{m=1}^{M}\alpha_m\cdot \big(\,u\,\mathbb{I}\,[\,u < u_m\,] + u_m\,\mathbb{I}\,[\,u\ge u_m\,] -  u\,u_m/\ell\,\big) \\
&+ \sum_{l=1}^{L}\beta_l\cdot \Big( \frac{u}{\ell}\int_{0}^{\ell}\int_{0}^{s}z'_l(v,x(v))dv\,ds -
\int_{0}^{u}\int_{0}^{s}z'_l(v,x(v))dv\,ds\,\Big)
\\
&+ \sum_{j=1}^{J}\gamma_j \cdot \Big( \frac{u}{\ell}\int_{0}^{\ell}\int_{0}^{s}dq_j(v) -
\int_{0}^{u}\int_{0}^{s}dq_j(v)\,ds\Big)
\ , \quad\quad
\quad u\in[0,\ell]\ .
\end{align*}
\end{prop}
\begin{proof}
See the Appendix.
\end{proof}
\noindent Thus, in both cases the  first terms appearing in the specification of $\{\Cov^{bm}\delta\Phi(x)\}$, $\{\Cov^{bb}\delta\Phi(x)\}$ in the proposition, are
continuous and piece-wise linear in $u$ (there is a turn at the time instances of the observations) so still lie within the Cameron-Martin spaces $\Hs_0^{bm}$,
$\Hs_0^{bb}$ respectively (even if the variational derivative $\delta \alpha$ itself will not necessarily lie within the Hilbert space, as shown in the
proof). The second terms are clearly a.s.\@ elements of the corresponding spaces $\Hs_0^{bm}$,
$\Hs_0^{bb}$ under weak continuity conditions on $z'_l$. Finally, for the third terms, again weak regulatory conditions on $r_j$ and $r'_j$ guarantee
that the corresponding paths in $u$ are elements of the appropriate Cameron-Martin spaces.

\subsection{Example Models}

\subsubsection*{Diffusions Observed at a Discrete Skeleton}

\noindent We first consider the case of discretely observed diffusions.
In other words, suppose that the diffusion $V$ is observed at times
$\{u_i, \;i=0,1,\dots, M\}$, with $u_0=0$ and $u_M=\ell$, and the observations are denoted with:
$$
Y=\{Y_i=V_{u_i}, \;i=0,1,\dots,m\}\ .
$$
The observations on the transformed unit-volatility diffusion $X$, defined in (\ref{eq:diffusionunitvol}), are then:
$$
\dot{Y}=\{\dot{Y}_i=X_{u_i}=\eta(Y_i;\theta), \;i=0,1,\dots, M\}\ .
$$
The reparametrisation of \cite{rob:str01} involves another transformation that operates on each of the
 $m$ independent constituent bridges.
Between observations $u_{i}$ and $u_{i+1}$ we define $\tilde{X}_u$ as
$$
\tilde{X}_u=X_u- \frac{(u_{i+1}-u)\dot{Y}_i+(u-u_{i})\dot{Y}_{i+1}}{u_{i+1}-u_{i}}, \quad u_{i}\leq u \leq u_{i+1}\ ,
$$
so that $X_u= b(\tilde{X}_u;\theta)$ with $b(\cdot;\theta)$ defined as the inverse of the above linear mapping.
For ease of exposition, we further assume that an antiderivative of $\nu(\cdot ;\theta)$ exists.
Then, we can write the target log-density $\log((d\Pi/d\Pi_0)(\tilde{x}|\dot{y}_i,\dot{y}_{i+1},\theta))$
for the path $\tilde{X}$ between $u_{i}$ and $u_{i+1}$ as (up to an additive normalising constant):
\begin{equation}
\label{eq:density discretely observed}
-\tfrac{1}{2}\int_{u_i}^{u_{i+1}}\left[ \nu{'}\big( b(\tilde{x}(s);\theta);\theta\big)+
\nu^2\big(b(\tilde{x}(s);\theta);\theta\big)\right]ds\ ,
\end{equation}
where $\nu{'}(x;\theta)$ denotes the derivative of $\nu(x ;\theta)$ with respect to
$x$ and $\Pi_0$ is the distribution of a standard Brownian bridge (note we used It\^o's Lemma in
(\ref{eq:density discretely observed})
to obtain an equivalent expression for the log-density that does not involve stochastic integral,
though this is not a requirement for our samplers).

\subsubsection*{Diffusions observed with error}

\noindent In this observation regime the data form a discrete skeleton of the diffusion path
in the presence of measurement error. $V$ is observed at the time instances $\{u_i, \;i=1,\dots, M\}$,
with  $u_M=\ell$, with observations  $Y=\{Y_i, \;i=1,\dots, M\}$. We consider the case of independent
observations conditional on the diffusion process and work with the unit volatility diffusion $X$,
such that $V_u=\eta^{-1}(X_u;\theta)$.
The probability density of the data given the latent path $X$ is:
\begin{equation}
\label{eq:diffusionserror}
p\big(y|\eta^{-1}(x;\theta);\theta\big)=\prod_{i=1}^M f\left(y_i|\eta^{-1}(x({u_i});\theta);\theta\right)\ ,
\end{equation}
where $f$ denotes the likelihood of the data given $V,\theta$.
From, (\ref{eq:diffusionserror}),  (\ref{eq:targetGenDiff})
the target log-density $\log((d\Pi/d\Pi_0)(x|y,\theta))$ is (up to an additive normalising constant):
\begin{align*}
\sum_{i=1}^{M}\Big\{\log f\big(y_i|\eta^{-1}(x({u_i});\theta);\theta\big) + \int_{u_i}^{u_{i+1}} \nu(x(s);\theta)dx(s)-
\frac{1}{2}\int_{u_i}^{u_{i+1}} \nu^2(x(s);\theta)ds\Big\}
\end{align*}
where now $\Pi_0$ denotes the distribution of a standard Brownian motion.

\subsubsection*{Stochastic volatility models}

\noindent Stochastic volatility models are  popular in modelling financial and econometric time
 series and instruments such as asset prices, interest rates and financial derivatives.
In standard stochastic volatility models for asset prices, such as those in \cite{hul:whi87} and \cite{hes93},
the log-price $S$ is a diffusion whose volatility is driven by another diffusion.
The SDE of the bivariate diffusion model has the following form
\begin{equation}
\label{eq:sv}
\begin{cases}
  dS_u=\mu_s(V_u;\theta)du+ \rho\sigma_s(V_u;\theta)dW_u+\sqrt{1-\rho^2}\,\sigma_s(V_u;\theta)dB_u\ ;\\
  dV_u=\mu_v(V_u;\theta)du+\sigma_v(V_u;\theta)dW_u\ ,
\end{cases}
\end{equation}
where $B$ and $W$ are independent Brownian motions and $\rho$ reflects the leverage effect.
For ease of exposition, we set $\rho=0$. Stochastic volatility models with leverage effect are still
into the framework of this paper and can be handled with the methodology described in Section \ref{sec:GeneralDiff}.
The Lamperti transform can be applied on $V$ to obtain the unit volatility diffusion $X$ as
in (\ref{eq:diffusionunitvol}).
Consider a pair of observations ($y_0$, $y_1$) denoting the value of $S$ at times $0$, $u_1$ respectively.
The distribution of the data conditional on the path of $X$ from $0$ to $u_1$ has a known closed form and we can write:
\begin{equation}
\label{eq:svlikelihood}
\begin{cases}
  y_1|y_0,x\;\sim\; N\left(M_{y}(x;\theta), V_{y}(x;\theta)\right)\ ;\\
  M_{y}(x;\theta)=y_0+\int_{0}^{u_1}\mu_s(v(u);\theta)du\ ;\\
  V_{y}(x;\theta)=\int_{0}^{u_1}\sigma_s(v(u);\theta)^2du\ ;\\
  x(u)=\eta^{-1}(v(u);\theta)\ ;\\
  dX_u=\nu(X_u;\theta)du+dW_u\ , \quad 0\leq u\leq u_1\ .\;
\end{cases}
\end{equation}
Cases with more observations are handled by splitting the data into pairs of consecutive points as above
and then, using the Markov property for the bivariate diffusion $(S,X)$, multiplying
the corresponding densities calculated according to (\ref{eq:svlikelihood}). The log-density $\log((d\Pi/d\Pi_0)(x|y,\theta))$
for the latent diffusion $X$ that drives the volatility for $0<u\leq u_1$ is (up to an additive normalised constant):
\begin{align*}
-\tfrac{\log V_{y}(x;\theta)}{2} -
 \tfrac{\left[y_1-M_{y}(x;\theta)\right]^2}{2V_{y}(x;\theta)}+
\int_{0}^{u_{1}} \nu(x(s);\theta)dx(s)-\frac{1}{2}\int_{0}^{u_{1}} \nu^2(x(s);\theta)ds \ ,
\end{align*}
where $\Pi_0$ denotes here standard Brownian motion.

\subsubsection*{Latent Diffusion Survival models}

\noindent Survival models target the probability of an individual $i$ surviving up to time
$u$ or else $P(Y>u)$, where $Y$ denotes the event time.
The aim is to model the hazard function $h(u)$ that reflects the
probability that an event will occur in the infinitesimal period $[u,u+du)$
retrieving information from available data in the form of event times.
Latent diffusion survival models \citep{aal:gje04,rob:san10} provide parametric formulations for $h(u)$,
which is assumed to be a positive function $h(\cdot)$ of a diffusion process.
The motivation in such models is to consider an underlying process that results in the
occurrence of each event \citep{aal:gje04}. The distribution function for a single observation $y_i$ is given by
$$
F\big(y_i|x,\theta\big)=1-\exp\big(-\int_{0}^{y_i} h(x(s);\theta)ds\big)\ ,\quad 0<y_i\leq \ell\ ,
$$
with density
$$
f\big(y_i|x,\theta\big)=h(x({y_i});\theta)\exp\Big(-\int_0^{y_i}h(x(s);\theta)ds\Big)\ ,\quad 0<y_i\leq \ell\ ,
$$
where $x(u)$ obeys a diffusion process. For ease of exposition we assign the unit diffusion defined by
 (\ref{eq:diffusionunitvol}) although other choices are possible.
The likelihood for the observed event times $y=(y_1,\dots,y_n)$, with $\max_i y_i \leq \ell$, can be written as:
\begin{equation}
\label{eq:diffusionssurvival}
f\big(y|x,\theta\big)=\Big[\prod_{i=1}^n h(x({y_i});\theta)\Big]\exp\Big(-\sum_{i=1}^n\int_0^{y_i}h(x(s);\theta)ds\Big)\ .
\end{equation}
Hence, the log-density $\log((d\Pi/d\Pi_0)(x|y,\theta))$ for the latent
diffusion $X$ becomes (up to an additive normalising constant):
\begin{align*}
\sum_{i=1}^n\Big\{\log h\left(x({y_i});\theta\right)-&\int_0^{y_i}h(x(s);\theta)ds\Big\}\\
&+\int_{0}^{\ell} \nu(x(s);\theta)dx(s)-\frac{1}{2}\int_{0}^{\ell} \nu^2(x(s);\theta)ds
\end{align*}
with $\Pi_0$ denoting the distribution of a standard Brownian motion as before. For more information about such models, including cases of censored data, the reader is referred to \cite{rob:san10}.

\section{Numerical Applications}
\label{sec:Numerical}

In this section, we employ the algorithms of Table \ref{tab:MCMC} to perform various simulation experiments involving diffusion bridges, stochastic volatility and latent diffusion survival models. In all these experiments, we treat the parameter vector $\theta$ as known and focus on the update of the latent diffusion path. The aim is to assess and compare the performance of the algorithms on various aspects including efficiency of the MCMC output and central processor unit (CPU) time. In order to measure CPU time in two different computing environments, the simulations for diffusion bridges and stochastic volatility models were carried out in MATLAB, whereas for the latent diffusion survival models the \texttt{C} programming language was used.

The draws of each algorithm are given by a Markov chain with the target posterior as equilibrium marginal distribution. After a suitable burn-in period, inference is based on these sequences that exhibit an amount of serial correlation and are therefore not as efficient as a posterior sampler with i.i.d. draws. This dependence can be measured by the so-called inefficiency factor, or else autocorrelation time
$$
INF(K)=1+2\sum_{k=1}^K\gamma(k),
$$
where $\gamma(k)$, is the autocorrelation at lag $k$ and $K$ is a suitably chosen truncation point. Loosely speaking the MCMC sampler must be run INF(K) times as many iterations, after the burn-in period, to match the variance of the posterior estimate obtained from a hypothetical independent posterior sampler. In this paper we focus on the inverse of INF(K), known as the effective sample size (ESS), to measure the relative efficiency of the proposed MCMC samplers. The use of INF(K) or ESS(K) is frequent in the MCMC literature; see e.g. \cite{geye:92} and, more specific to the context of this paper, \cite{ch:pit:she06} and \cite{giro:cald11} for applications on stochastic volatility models and Hybrid Monte Carlo respectively. In \cite{giro:cald11}, in order to summarize the inefficiencies for all the model parameters, the minimum ESS(K) was used that was taken over a number of the univariate MCMC trajectories. In our context, the MCMC performance is assessed by monitoring the posterior draws of the diffusion, recorded at a fine partition of its path, and reporting the minimum ESS(K) over these points. By repeated applications of the algorithm, we set both $K$ and the number of iterations of the MCMC algorithm to high enough values so that the minimum ESS(K) stabilizes for all algorithms. We therefore suppress the notation to ESS for the remainder of the paper. The value of ESS was multiplied by a factor of $100$ to reflect the percentage of the total MCMC iterations that can be considered as independent draws from the posterior.

The employed MCMC algorithms consist of an independence sampler proposing from the reference Brownian path $\Pi_0$ and the advanced algorithms in Table \ref{tab:MCMC}. The algorithms were tuned to achieve certain acceptance probability levels that, according to our experience and previous literature, are associated with better performance. Specifically we aimed in attaining an acceptance probability around ($15\%$ - $30\%$) for RWM, ($50\%$-$70\%$) for MALA and ($65\%$-$85\%$) for HMC. To explore the performance of  HMC we first fixed the number of leapfrog steps (e.g. to 5 or 10) and then recorded the minimum ESS for various levels of acceptance probability. We then considered cases with additional leapfrog steps. For each of these algorithms, we monitor the values of the minimum ESS, CPU times and their ratio in absolute and relative terms. The presented results contain the best version of these algorithms.

\subsection{Diffusions Observed at a Discrete Skeleton}
\noindent Consider the diffusion  discussed in Section \ref{sec:OUanalytical}, i.e. an OU process with SDE:
$$
dX_u=-\kappa X_u du +dB_u\ , \quad  0 \leq u \leq \ell\ ,
$$
with $X_0=0$ and an observation at time $\ell=1$. We set $X_1=0$ and consider 3 different values for
$\kappa$, i.e. $12,20,30$ in our investigation of the MCMC performance. The algorithms were constructed using the Euler-Maruyama approximation of the target density (\ref{eq:density discretely observed}) from a time-discretised diffusion path. The MCMC components comprise of the equidistant points from a discrete skeleton of the diffusion. The discretisation step was set to $\delta=0.02$. Table \ref{tab:sim1} provides the results, i.e. the values of the minimum ESS, CPU times and their absolute and relative ratio. The HMC algorithm consisted of $5$ leapfrog steps with the parameter $h$ set to values $(0.43,0.26,0.17)$ for values of $\kappa$ $(12,20,30)$ respectively. For advanced MALA, that can be thought as HMC with a single leapfrog step, the corresponding values of $h=\sqrt{\Delta t}$ were very similar $(0.45,0.26,0.18)$ indicating much smaller total steps. Overall, advanced HMC consistently over-performs, in terms of ESS, the remaining algorithms. In particular for $\kappa=30$,  HMC  is faster than the independence sampler by a factor of over $30$. Its performance remains at high levels as we increase $\kappa$ and does not deteriorate as $\delta$ becomes smaller, as indicated by the results obtained for $\delta=0.01$ and $\delta=0.005$. In line with the results of \cite{whit:09} and Section \ref{sec:OUanalytical}, we note a substantial improvement over advanced MALA suggesting a more efficient use of the gradient within HMC. MALA offers some improvement over RWM and independence sampler, but at a heavy additional computational cost. The independence sampler performs reasonably well for $\kappa=12$ (acceptance rate of $16\%$) but its performance drops substantially as $\kappa$ increases and the acceptance rate becomes smaller; $8\%$ for $\kappa=20$ and $1.2\%$ for $\kappa=30$.

\begin{table}[!h]
\begin{tabular}{lcccc}
\hline
$\kappa=12$ & min(ESS) & time &  $\frac{\min(\textrm{ESS})}{\textrm{time}}$ \vspace{0.1cm} & relative $\frac{\min(\textrm{ESS})}{\textrm{time}}$\\
\hline
IS &  3.9173  &	  4.8811   &	0.8025	  &  2.1733  \\
RWM                  &  3.9584  &     5.8925   &	0.6718	  &  1.8192  \\
MALA                 &  4.0112  &    10.8626   &	0.3693	  &  1 \\
HMC                  & 35.7274	&    20.8695   &    1.7119	  &  4.6361 \\
HMC ($\delta=0.01$)  & 35.8903  &    32.5594   &    N/A   &  N/A\\
HMC ($\delta=0.005$) & 35.5875  &    51.6085   &    N/A   &  N/A\\
\hline
$\kappa=20$	& min(ESS) & time  & $\frac{\min(\textrm{ESS})}{\textrm{time}}$ \vspace{0.1cm} &  relative $\frac{\min(\textrm{ESS})}{\textrm{time}}$\\
\hline
IS &  0.5013	&  4.4977	& 0.1115	&  1        \\
RWM                  &  1.0086	&  5.4445	& 0.1853	&  1.6621   \\
MALA                 &  1.6202	& 10.0588	& 0.1611	&  1.4452   \\
HMC                  & 26.6214	& 20.8841	& 1.2747	& 11.4369   \\
\hline
$\kappa=30$	& min(ESS) & time & $\frac{\min(\textrm{ESS})}{\textrm{time}}$ \vspace{0.1cm}  & relative $\frac{\min(\textrm{ESS})}{\textrm{time}}$\\
\hline
IS &  0.1012	 &  4.7043	&  0.0215	&  1  \\
RWM                  &  0.4343	 &  5.7229	&  0.0759	&  3.5277 \\
MALA                 &  0.5372	 & 10.0438	&  0.0535	&  2.4863 \\
HMC                  & 13.3350	 & 20.4831	&  0.6510	& 30.2631\\
\hline
\end{tabular}
\caption{Relative efficiency via the minimum ESS (\%) and CPU times (seconds), for the advanced pathspace algorithms - Case of OU bridges. IS denotes the Independence Sampler.}
\label{tab:sim1}
\end{table}

\subsection{Stochastic Volatility Models}
\noindent The following stochastic volatility model was used to simulate data:
\begin{equation*}
\begin{cases}
  dS_u=\exp(V_u/2)dB_u\ ,\quad 0\leq u\leq \ell\ ;\\
  dV_u=\kappa(\mu-V_u)du+\sigma dW_u\ .
\end{cases}
\end{equation*}
The parameters were set according to previous analyses based on similar models for the S\&P 500 dataset
\citep{ch:pit:she06}. Specifically, we set $\kappa=0.03$, $\mu=0.07$, $\sigma^2=0.03$ and $V_0=0$.
We consider about a year measured in days ($\ell=250$) and recorded observations at a daily frequency ($250$ data points).
The transformation of $V_u$ to a unit volatility diffusion was used to write the target density and construct the HMC algorithms.
The model for a pair of consecutive observations, ($y_{i-1},y_{i}$) can be written as:
\begin{equation*}
\label{eq:svlikelihoodex}
\begin{cases}
  y_{i}|y_{i-1}\;\sim\; N\big(y_{i-1},\; \int_{u_{i-1}}^{u_{i}}\exp(\sigma x(s))ds\big)\  ;\\
  x(u)=v(u)/\sigma\ ; \\
  dX_u=\kappa\left(\frac{\mu}{\sigma}-X_u\right)du+dW_u\ , \quad u_{i-1}\leq u\leq u_i\ .
\end{cases}
\end{equation*}
The results are shown in Table \ref{tab:sim2}. Independence sampler performs very poorly in this case, with an
acceptance rate below $10^{-4}$, and is omitted from the table. MALA provides a small improvement over RWM which is nevertheless not enough to cover the associated increase in the corresponding computations. However, this is not the case for HMC that reaches its optimal performance roughly at $10$ leapfrog steps. Advanced HMC  offers considerable improvement, being nearly $8$ times faster than RWM and 11 times faster than MALA. Parameter $h$ that corresponded to the desired acceptance probability levels was $0.085$ for  MALA algorithm and $0.075$ for all the versions of  HMC.

\begin{table}
\begin{tabular}{lcccc}
\hline
Sampler  & min(ESS) & time &  100$\times$ $\frac{\min(\textrm{ESS})}{\textrm{time}}$ \vspace{0.1cm} & relative $\frac{\min(\textrm{ESS})}{\textrm{time}}$\\
\hline
RWM                       & 0.1400	  &  161.8298	&  0.0865	&  1.3561    \\
MALA                      & 0.2181	  &  341.8737	&  0.0638	&  1.0000   \\
HMC ($5$ steps)           & 2.5695	  &  689.6767	&  0.3726	&  5.8400   \\
HMC ($10$ steps)          & 8.1655	  & 1188.1201	&  0.6873	& 10.7729   \\
HMC ($20$ steps)          &  8.3216   & 2200.1311	&  0.3782	&  5.9288   \\
\hline
\end{tabular}
\caption{Relative efficiency, via the minimum ESS (\%) and CPU times (seconds) for the diffusion pathspace algorithms - Case of stochastic volatility paths.}
\label{tab:sim2}
\end{table}

\subsection{Latent Diffusion Survival models}
\begin{figure}
\begin{flushleft}
\includegraphics[scale=0.5]{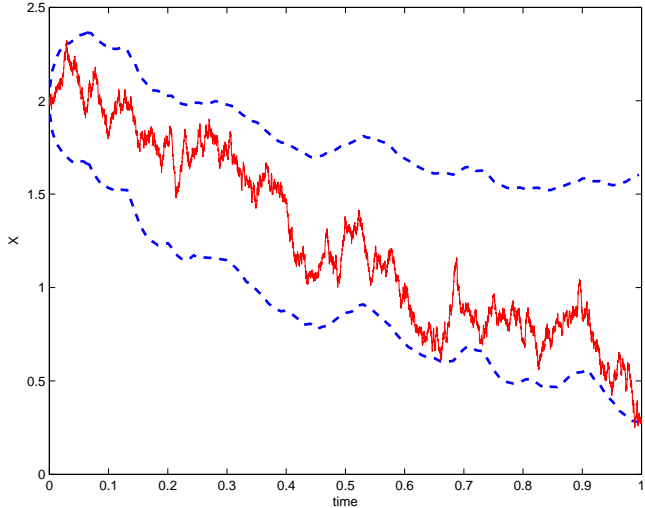}
\caption{$95\%$ Pointwise credible intervals (blue dashed lines) overlayed on the true path of $X$ (red solid line). }
\label{fig:sim1}
\end{flushleft}
\end{figure}

\noindent This section provides a numerical illustration in simulated data from a latent diffusion survival
 model appearing in \cite{rob:san10}. Specifically, the likelihood for event times $y=\{y_1,\dots,y_n\}$ is given by:
\begin{equation*}
p\big(y|\eta^{-1}(x;\theta);\theta\big)=\Big[\prod_{i=1}^n x^2({y_i})\Big]\exp\Big(-\sum_{i=1}^n\int_0^{y_i}x^2(s)ds\Big)
\end{equation*}
where we have considered the model:
\begin{equation}
\label{eq:SDEsimSurv}
dX_u=-(1.4\sin(X_u)+1)du+dB_u\ ,\quad X_0=2\ .
\end{equation}
We simulated a trajectory driving the hazard function from \eqref{eq:SDEsimSurv} that
can be seen in Figure \ref{fig:sim1} (solid red line). Despite various fluctuations,
 the simulated hazard is decreasing and could be used, for example, in cases of lifetime data
from patients following a surgical operation. The trajectory of Figure \ref{fig:sim1} was then used,
 through the distribution function it defines, to draw $200$ i.i.d.\@ event times.
In the simulation exercise of this section, we treat the parameters of the diffusion as known and apply the various MCMC algorithms of Table \ref{tab:MCMC} to sample from the posterior of the latent diffusion process.
The SDE in \eqref{eq:SDEsimSurv} can also be thought as a prior for the hazard function trajectory.
Our focus is the behaviour and efficiency of the proposed MCMC algorithms,
while keeping an eye on the ability of the model to capture the shape of the unobserved hazard.

Table \ref{tab:sim3} provides the measures of performance of the algorithms in Table \ref{tab:MCMC}.
The calculations in this section were obtained using \texttt{C} programming language, unlike the
previous two applications where MATLAB was used. Similarly with the stochastic volatility simulation experiment,
the independence sampler is associated with extremely low acceptance rate rendering it infeasible in practice.
RWM also performs poorly. A very small step is required to achieve the desired acceptance rate,
thus resulting in very small moves around the diffusion pathspace. MALA with $h=0.2$ performs better
in this case, but massively better performance is offered by advanced HMC. Specifically, HMC with $10$
 leapfrog steps and $h=0.15$ is about $54$ times faster than RWM. As already mentioned,
Figure \ref{fig:sim1} depicts the trajectory of the latent diffusion process that determines
the hazard function and was used to generate the data. It also displays $95\%$ pointwise credible
intervals, obtained from the HMC algorithm, which indicate that the shape of the hazard function is captured reasonably well.

\begin{table}
\begin{tabular}{lcccc}
\hline
Sampler  & min(ESS) & time &  100$\times$ $\frac{\min(\textrm{ESS})}{\textrm{time}}$ \vspace{0.1cm} & relative $\frac{\min(\textrm{ESS})}{\textrm{time}}$\\
\hline
RWM                       & 0.1039	  &   55.2342	&  0.1881   &	1      \\
MALA                      & 0.6466    &	  87.5021	&  0.7389	&   3.9284 \\
HMC                       & 25.2985	  &  248.0301	& 10.1997	&  54.2229 \\
\hline
\end{tabular}
\caption{Relative efficiency, via the minimum ESS (\%) and CPU times (seconds) for the advanced  pathspace algorithms - Case of latent diffusion survival model.}
\label{tab:sim3}
\end{table}

\section{Robustness in Dimension in Further Scenarios}
\label{sec:GeneralDiff}

We will briefly illustrate the well-definition of advanced HMC algorithms for cases when the target
distribution of the SDE-driven models can have a different structure from the ones considered so far.
In what follows, we omit reference to the parameter $\theta$.

\subsection{Beyond the Lamperti Transform}
In the context of a non-scalar diffusion $V_u$, defined via the equation (\ref{eq:diffusion}),
it is not guaranteed at all that $V$ can be transformed into an SDE of unit diffusion coefficient.
Indeed, the Lamperti transform in such a context would require the existence of a mapping $V_u\mapsto \eta(X_u)$ (with $\eta = (\eta_1,\eta_2,\ldots, \eta_d)^{\top}: \R^{d}\mapsto \R^{d}$) such that, for all $v$ in the state space of $V_t$:
\begin{equation}
\label{eq:Laberti}
 D\eta(v)\cdot \sigma(v) = I_{d}
\end{equation}
where $D\eta  = ( \partial \eta_i / \partial v_j )$. This follows directly by It\^o's formula, see e.g.\@   \cite{saha:08}.
The work in \cite{saha:08} also shows that existence of a mapping $\eta$ with the property (\ref{eq:Laberti}) is equivalent to the diffusion coefficient matrix
satisfying $\partial \sigma^{-1}_{ij}/\partial v_k = \partial \sigma_{ik}/\partial v_j$ for all $i,j,k$ with $j<k$. This  restricts considerably the
applicability of the Lamperti transform for non-scalar diffusions. There are however other methods suggested in the literature of a wider scope.
Our advanced MCMC algorithms require posterior distributions on pathspace which are absolutely continuous w.r.t.\@ Brownian motion-related
distributions, so it is of interest to briefly verify their well-definition on the pathspace when using such alternative transforms.

The method  in \cite{ch:pit:she06} maps $V_u$ onto the driving Wiener noise $X_u=B_u$
of the SDE. Assuming some data $Y$ with likelihood
$p(y|v)$, and since the prior on $X$ is simply the Wiener measure $\Pi_0 = N(0,\Cov^{bm})$,
we can write the posterior on $X$ as:
\begin{equation*}
 \frac{d\Pi}{d\Pi_0}(x) \propto p(y|v) \ .
\end{equation*}
(Note, that we use here the mapping $x\mapsto v$ between driving noise $x$ and process $v$
implied by the expression of the SDE for the process $V$.)
It remains to calculate  $\Cov^{bm}\, \delta \Phi(x)$ is this context.
Differentiation of $\Phi(x)$ will involve finding derivatives of $v$ w.r.t.\@
the driving noise $x$.
So it is not a surprise that the dynamics of the  Malliavin derivative $D_s V_u$ (see e.g.\@ \cite{four:99})
will appear in the calculations;
this is defined as below:
\begin{gather*}
\tfrac{dY_u}{Y_u} = \mu'(V_u)du+ \sigma'(V_u)\,dB_u \ , \quad Y_0 = 1 \ ;  \\
D_s V_u = \tfrac{Y_u}{Y_s}\,\sigma(V_s)\,\mathbb{I}\,[\,s\le u\,]\ .
\end{gather*}
We will assume here the general structure for $\Phi(x)=-\log(p(y|v))$ (compared with the structure assumed earlier in (\ref{eq:structure}) we do not
include here stochastic integral terms to avoid overly cumbersome expressions):
\begin{align}
\nonumber
\Phi(x)  = &\alpha(v(u_1),v(u_2),\ldots, v(u_M)) \\ &+
\beta\big( \int_{0}^{\ell}z_1(s,v(s))ds,\,\int_{0}^{\ell}z_2(s,v(s))ds,\, \int_{0}^{\ell}z_L(s,v(s))ds\big)
 \label{eq:structure2} \ .
\end{align}
The terms $\alpha_m$, $\beta_l$ appearing below correspond to partial derivatives
of the functionals $\alpha$, $\beta$ as in the case of Proposition \ref{prop:3}.
\begin{prop}
\label{prop:transform}
For the functional $\Phi(x)$ given in (\ref{eq:structure2}), for any
$x\in\Hs$:
\begin{align*}
\big(\Cov^{bm}\delta\Phi(x)\big)&(u) =
\sum_{m=1}^{M}\alpha_m\cdot\Big(\,(u\wedge u_m)\,(F_{m,u_m} + \sigma(v({u_m}))) - \int_{0}^{u\wedge u_m}F_{m,r} dr\,\Big)\\
& +  \sum_{l=1}^{L} \beta_l\cdot \,\Big(\,u\,( G_{l,\ell} + J_{l,\ell})+ \int_{0}^{u}( G_{l,r} + J_{l,r} ) dr\Big) \ ,
\quad u\in[0,\ell] \ ,
\end{align*}
for the processes, for $m=1,2,\ldots, M$ and $l=1,2,\ldots, L$:
\begin{gather*}
F_{m,r} = \int_{0}^{r} e^{\int_{s}^{u_m}\big(\mu'(v(u))du + \sigma'(v(u))dx(u)\big)}\,dQ_s \ ;\\
G_{l,r} = \int_0^{r} \int_{s}^{\ell} z'_l(t,v(t))\,e^{\int_{s}^{t}\big(\mu'(v(u))du + \sigma'(v(u))dx(u)\big)}dt\,dQ_s\ ; \\
J_{l,r} = \int_{0}^{r} z_l'(s,v(s))\,\sigma(v(s))ds \ ,
\end{gather*}
with integrator:
\begin{equation*}
dQ_s = \sigma(v(s))( \mu'(v(s))ds + \sigma'(v(s))dx(s) ) - d\sigma(v(s))\ .
\end{equation*}
\end{prop}

Focusing on the properties of the calculated path $\Cov^{bm}\delta\Phi(x)$  over its domain $u\in [0,\ell]$, it is easy to check the following: a.s.\@ in  $x$, the first terms are continuous, piecewise linear with
a turn at the data instances $u_1, u_2, \ldots, u_M$; then, under the weak assumption
that the processes $r\mapsto G_{l,r}$, $r\mapsto J_{l,r}$ are a.s.\@  continuous, we have that the second terms in the calculation in Proposition \ref{prop:transform}
are a.s.\@ differentiable. Thus, under weak conditions Assumption \ref{ass:ass} is satisfied and advanced HMC is well-defined on the pathspace.

\section{Discussion - Further Directions}
\label{sec:Discussion}

In this paper we developed and applied a general class of derivative driven algorithms, suitable for Hilbert spaces defined by diffusion driven models. The validity of the algorithm was established using different theory than \cite{besk:11}, which allowed us to relax involved assumptions and generalize to widely used diffusion models across various scientific fields as mentioned in Section \ref{sec:introdif}. The modelling framework contains diffusions observed with error, partial data, as well as observation on functionals of the diffusion process such as hitting times and integrals thereof. The extended framework also includes diffusions processes with general diffusion coefficients that can be handled with reparametrisation to the driving Wiener noise as in \cite{ch:pit:she06} and \cite{goli:08}.
A number of research directions related with the methods presented in this paper have remained open:

\begin{itemize}
\item The algorithm can be used in contexts where a Gibbs data augmentation scheme is adopted to facilitate the step of updating the diffusion path given the parameters. The entire diffusion path can either be updated in a single step, but can also be integrated to updates based on overlapping and potentially random-sized blocks. Note that while this approach will boost the conditional step of the diffusion path updates,
another important aspect for the overall performance of the MCMC sample is the posterior correlations between diffusion paths and parameters. The role of reparametrisation of diffusion paths is critical for this task. We are currently investigating extensions of the advanced  HMC algorithm on the joint density of diffusion and parameters. In the numerical applications of this paper the high dimensional diffusion path was successfully updated in a single block, a fact that provides supporting evidence that the algorithm could still be very efficient when also incorporating the low dimension parameter vector $\theta$ in it. Such an approach will provide an alternative to pseudo marginal \citep{and:rob09} or particle MCMC \citep{and:dou:hol10} formulations for diffusion driven models; see e.g.\@  \cite{str:bog11} and \cite{dur:kal:bag12} respectively.
\item In this paper the choice of the mass matrix for the Hamiltonian dynamics is guided by the prior, i.e.\@ by the reference Gaussian law $N(0,\Cov)$.
Recent work in the literature (\cite{giro:11}) has looked at exploiting the geometric structure of state-space to consider location-specific mass matrices.
It is therefore worth investigating the choice of location-specific mass matrices in the context of infinite-dimensional pathspaces to further boost
the efficiency of MCMC methods.
\item An important direction of further research is related with the calibration of long-memory diffusion models: an important class
covers the case of processes driven by fractional Brownian motion (see e.g.\@ \cite{tudo:07, kou:08}). Applying standard Independence samplers
over smaller blocks of the complete latent path will be overly expensive as calculations have to be done every time
 over the whole path due to the lack of Markovianity. In this case it could be very beneficial to update the whole path together
using HMC and control the acceptance probability by trying different leapfrog steps.
\end{itemize}

\section*{Acknowledgments}
We are grateful to two referees for many comments that greatly improved the appearance of the paper.
We also thank Dr Yvo Pokern for a lot of useful advices during the development of the paper.
EP acknowledges support from EPSRC for his PhD studies.
\appendix
\section{Proofs}
\begin{proof}[Proof of Proposition \ref{prop:1}]
We  use the notation $|\cdot|_{L_1}\equiv \Exp\,|\cdot|$.
For both algorithms (RWM and MALA) the acceptance probability can be expressed as:
\begin{equation*}
a(x,v) = 1 \wedge e^{R(x,v)}
\end{equation*}
for the relevant variable $R(x,v)$.
Working as in \cite{whit:09}, for the first result it suffices to prove that
$\sup_{\ell}|R(x,v)|_{L_1} < \infty$  since we have the inequality:
\begin{equation*}
\Exp\,[\,1\wedge e^{R}\,] \ge \Exp\,[\,e^{-|R|}\,] \ge e^{-|R|_{L_1}}\ .
\end{equation*}
(We thank a referee for pointing at the above inequality, which replaced a slightly more complex expression
we originally had.)
For the second result,
 we will be identifying a term in $R$ with the highest order $L_1$-norm, which we will
denote by $J$. Crucially, this term will lie in $(-\infty,0)$, thus will be growing to $-\infty$
faster than the growth of  $|R-J|_{L_1}$. This provides some intuition for the zero limit of the
average acceptance probability. Analytically, as in \cite{whit:09}, we will be using the following  inequality,
 for any $\gamma>0$:
\begin{align}
\Exp[\,1&\wedge e^{R}\,] \le \mathrm{P}[\,R\ge -\gamma\,] + e^{-\gamma}  \nonumber \\
 & = \mathrm{P}\,[\,\{ R \ge -\gamma \} \cap \{ |R-J|\le \gamma  \} \,] +
 \mathrm{P}\,[\,\{ R \ge -\gamma \} \cap \{ |R-J| >\gamma  \} \,]
 + e^{-\gamma} \nonumber \\
 &\le \mathrm{P}\,[\,J\ge - 2\gamma\,] + \mathrm{P}\,[\,|R-J| > \gamma\,] + e^{-\gamma} \nonumber \\
 & \le \mathrm{P}\,[\,J\ge - 2\gamma\,] + \tfrac{|R-J|_{L_1}}{\gamma} + e^{-\gamma}\ ,
\label{eq:trick}
\end{align}
where we used the Markov inequality in the last line. So, the idea here will be to choose some $\gamma$
growing to $\infty$ faster that the rate of growth of $|R-J|_{L_1}$ and slower than the rate of growth
of $|J|_{L_1}$.

From \cite{whit:09}, we retrieve the following collection of results:
\begin{gather}
\Exp\,|x|^2 = \tfrac{\ell}{2\kappa \tanh(\kappa\ell)} - \tfrac{1}{2\kappa^2}\ ;\quad
\Exp\,|\Cov x|^2 = \tfrac{\ell^4}{90\kappa^2} - \tfrac{1}{2\kappa^6} - \tfrac{\ell^2}{6\kappa^4} +
\tfrac{\ell}{2\kappa^5\,\tanh(\kappa\ell)} \ ; \nonumber \\
\Exp\,[\,\langle x,\Cov x \rangle \,]  = \tfrac{3+\kappa^2\ell^2}{6\kappa^4} -
\tfrac{\ell}{2\kappa^3\tanh(\kappa\ell)}\ ;\\ \Exp\,[\,\langle x, \Cov^3 x \rangle \,] =
\tfrac{945+315\kappa^2\ell^2-21\kappa^4\ell^4+2\kappa^6\ell^6}{1890\kappa^8} -
\tfrac{\ell}{2\kappa^7\tanh(\kappa\ell)}\ ;\nonumber  \\
\Exp \,|v|^2  = \tfrac{\ell^2}{6}\  ;\quad
\Exp\,[\,\langle v, \Cov v \rangle \,] =\tfrac{\ell^4}{90}\ ;\nonumber \\
\Exp\,[\,\langle x, v \rangle^2 \,] = \tfrac{3+\kappa^2\ell^2}{6\kappa^4} - \tfrac{\ell}{2\kappa^3\tanh(\kappa\ell)}\ ;\quad
\Exp\,[\,\langle\Cov x, v \rangle^2\,] = \tfrac{945+315\kappa^2\ell^2-21\kappa^4\ell^4+2\kappa^6\ell^6}
{1890\kappa^8}\ ; \nonumber \\
\Exp\,[\,\langle\Cov^2 x, v \rangle^2\,] = \tfrac{467775+155925\kappa^2\ell^2-10395\kappa^4\ell^4+990\kappa^6\ell^6-99\kappa^8\ell^8+10\kappa^{10}\ell^{10}}{467775\kappa^{12}}\ . \label{eq:L1}
\end{gather}
%
%
For the case of RWM, analytical calculations will give:
\begin{equation*}
R(x,v) = \tfrac{\kappa^{2}}{2(1+\frac{\Delta t}{4})^2}\,\,\Delta t\, \langle x, x \rangle - \tfrac{\kappa^2}{2(1+\frac{\Delta t}{4})^2}\, \,\Delta t\, \langle v, v \rangle\  - \tfrac{\kappa^2(1-\frac{\Delta t}{4})}{(1+\frac{\Delta t}{4})^2}\, \,\sqrt{\Delta t}\, \langle x, v \rangle  \ .
\end{equation*}
Now, when $\Delta t = c/\ell^2$, from the results in (\ref{eq:L1}) we find that the $L_1$-norm of
 each of the three above summands is $\mathcal{O}(1)$, so $\sup_{l}|R(x,v)|_{L_1} < \infty$.
When $\Delta t = c/\ell^\epsilon$ with $\epsilon\in(0,2)$, a careful inspection of (\ref{eq:L1}) gives that
the term with the highest order $L_1$-norm is:
\begin{equation*}
 J = -\tfrac{\kappa^2}{2(1+\frac{\Delta t}{4})^2}\, \,\Delta t\, \langle v, v \rangle\ .
\end{equation*}
Indeed, we have that:
\begin{equation*}
 |J|_{L_1}= \Theta(\Delta t\,\ell^2)\ ;\quad
 |R-J|_{L_1} = \mathcal{O}(\sqrt{\Delta t}\,\ell)\ .
\end{equation*}
We will apply (\ref{eq:trick}) for the selection $\gamma = (\Delta t\,\ell^2)^{2/3}$.
Clearly, the last two terms in the R.H.S.\@ of (\ref{eq:trick}) vanish as $\ell\rightarrow \infty$.
For the first term we need to use the analytical definition of $J$. Using the rescaling properties
of the Brownian bridge,
we can write $v_{u\ell} = \sqrt{\ell}\,\tilde{v}_u$ for a standard Brownian bridge $\tilde{v}$ on~$[0,1]$.
Thus, we can re-write:
\begin{equation*}
\langle v, v \rangle = \int_{0}^{\ell} v^{2}_u du = \ell  \int_{0}^{1}v^{2}_{u\ell} du \equiv
\ell^{2}\int_{0}^{1}\tilde{v}^{2}_{u}\,du = \ell^2\,|\tilde{v}|^2\ .
\end{equation*}
We now have that:
\begin{equation*}
\mathrm{P}\,[\,J \ge -2\gamma\,] = \mathrm{P}\,[\,|\tilde{v}|^{2} \leq
\tfrac{4(1+\frac{\Delta t}{4})^2}{\kappa^2}\,\tfrac{\gamma}{\Delta t\,\ell^2}\,] \rightarrow
\mathrm{P}\,[\,|\tilde{v}|^{2} = 0\,] = 0 \ .
\end{equation*}
For the case of MALA, some tedious calculations detailed in \cite{whit:09} will give that:
\begin{align*}
R(&x,v) = \tfrac{\kappa^2}{8(1+\frac{\Delta t}{4})^2}\,\Delta t^2\,\langle x,x \rangle
 - \tfrac{\kappa^6}{32(1+\frac{\Delta t}{4})^2}\,\Delta t^3\,\langle \Cov x,\Cov x \rangle
- \tfrac{\kappa^4(1-\frac{\Delta t}{4})}{8(1+\frac{\Delta t}{4})^2}\,\Delta t^2\,\langle x,\Cov x \rangle  \\
&
+\tfrac{\kappa^2}{8(1+\frac{\Delta t}{4})^2}\,\Delta t^2 \langle x,\Cov x \rangle
- \tfrac{\kappa^6}{32(1+\frac{\Delta t}{4})^2}\,\Delta t^3\,\langle \Cov^2 x,\Cov x \rangle
+ \tfrac{\kappa^4(1-\frac{\Delta t}{4})}{8(1+\frac{\Delta t}{4})^2}\,\Delta t^2\,\langle \Cov x,\Cov x \rangle \\
&-\tfrac{\kappa^2}{8(1+\frac{\Delta t}{4})^2}\,\Delta t^2\,\langle v, v \rangle
-\tfrac{\kappa^2}{8(1+\frac{\Delta t}{4})^2}\,\Delta t^2\,\langle  v,\Cov v \rangle
-\tfrac{\kappa^2(1-\frac{\Delta t}{4})}{4(1+\frac{\Delta t}{4})^2}\,\Delta t^{3/2}\,\langle x, v \rangle  \\
&
+\tfrac{\kappa^4}{8(1+\frac{\Delta t}{4})^2}\,\Delta t^{5/2}\,\langle\Cov x, v \rangle
-\tfrac{\kappa^2(1-\frac{\Delta t}{4})}{4(1+\frac{\Delta t}{4})^2}\,\Delta t^{3/2}\,\langle\Cov x, v \rangle
+\tfrac{\kappa^4}{8(1+\frac{\Delta t}{4})^2}\,\Delta t^{5/2}\,\langle\Cov^2 x, v \rangle \ .
\end{align*}
Now, when $\Delta t = c/\ell^2$, from the results in (\ref{eq:L1}) we find that the $L_1$-norm of each of the twelve above summands is indeed $\mathcal{O}(1)$, so $\sup_{l}|R(x,v)|_{L_1} < \infty$.
When $\Delta t = c/\ell^\epsilon$ with $\epsilon\in(0,2)$, using again the results in (\ref{eq:L1}),
we identify, among the twelve terms in the definition of $R(x,v)$, the one with the highest $L_1$-norm:
\begin{equation*}
J = - \tfrac{\kappa^6}{32(1+\frac{\Delta t}{4})^2}\,\Delta t^3\,\langle \Cov^2 x,\Cov x \rangle  \ .
\end{equation*}
Indeed we have that:
\begin{equation*}
|J|_{L_1} = \Theta(\Delta t^3 \ell^6) \ ; \quad |R-J|_{L_1} = \mathcal{O}(\Delta t^{5/2} \ell^5)\ .
\end{equation*}
We note that the second result above comes from the bound on the $L_1$-norm of the last summand in the definition of $R(x,v)$. In this case we apply (\ref{eq:trick}) under the choice
$\gamma = (\Delta t\,\ell^2)^{11/4}$. As before, under this choice the last two terms on the R.H.S.\@ of (\ref{eq:trick}) vanish as $\ell\rightarrow \infty$. For, the first term we must use the analytical
definition of $J$ from above. To handle $J$ we can use the Karhunen-Lo\`eve expansion of Gaussian measures.
In particular, using the representation for an OU-bridge from (\ref{eq:KL}) we get that:
\begin{equation*}
\langle x, \Cov^{3}x \rangle = \sum_{p=1}^{\infty} \frac{1}{\frac{\pi^2 p^2}{\ell^2}+\kappa^2}\,
\frac{\ell^6}{\pi^6\,p^6}\,\,\xi_p^2\ .
\end{equation*}
Thus, we can now obtain:
\begin{align*}
\mathrm{P}\,[\,J\ge &-2\gamma\,] = \mathrm{P}\,\Big[\,\sum_{p=1}^{\infty}\tfrac{1}{\frac{\pi^2 p^2}{\ell^2}+\kappa^2}
\,\tfrac{1}{\pi^6 p^6}\,\xi_p^2 \le \tfrac{32(1+\frac{\Delta t}{4})^2}{\kappa^6}\,
\tfrac{1}{(\Delta t\ell^2)^{1/4}} \,\Big] \\
& \le
\mathrm{P}\,\big[\,\tfrac{1}{\pi^2+\kappa^2}
\,\tfrac{1}{\pi^6}\,\xi_1^2 \le \tfrac{32(1+\frac{\Delta t}{4})^2}{\kappa^6}\,
\tfrac{1}{(\Delta t\,\ell^2)^{1/4}} \,\big]\rightarrow \mathrm{P}\,[\,\xi_1^{2} \le 0\,] = 0\ .
\end{align*}
The proof is now complete.
\end{proof}

\begin{proof}[Proof of Proposition \ref{prop:2}]
We exploit the  representation of $\Psi_{h}^{I}$ in (\ref{eq:power}).
Recall that we denote $(x_i,v_i) = \Psi_h^{i}(x_0,v_0)$, for number of leapfrog steps $0\le i \le I$.
After tedious
calculations, one can obtain:
\begin{equation}
\label{eq:energy1}
 \Delta H(x_0,v_0) = H(x_{I},v_{I}) - H(x_0,v_0) \equiv
\langle \mathcal{A}x_0, x_0 \rangle + \langle \mathcal{B}v_0, v_0 \rangle +
 \langle \mathcal{G}x_0, v_0 \rangle
\end{equation}
for the operators:
\begin{gather}
\mathcal{A} = -\tfrac{1}{2}\,\sin^2(I\theta)\,\mathcal{P}\ ; \quad
\mathcal{B} = \tfrac{1}{2}\,\sin^2(I\theta)\,a^2\,\mathcal{P}\ ; \quad
\mathcal{G} = \tfrac{1}{2}\,\sin(2I\theta)\,a\,\mathcal{P}\ ; \nonumber \\
\mathcal{P} = \kappa^2\, I + (1-\tfrac{1}{a^2})\,(\Cov^{bb})^{-1}\ . \label{eq:define}
\end{gather}
We have used operator's notation: $\sin(I\theta)$ is the operator such that we have $\sin(I\theta)x = \{\sin(I\theta_p)\,x_p\}_{p=1}^{\infty}$
where $\{x_p\}$ are the co-ordinates of $x\in\Hs$ w.r.t.\@ to the orthonormal basis corresponding to the eigen-functions of $\Cov^{bb}$;
a similar interpretation stands for the operator $\sin(2I\theta)$. Also $a^{2}x = \{a^{2}_p\,x_p\}_{p=1}^{\infty}$.
The sequences $\{\theta_p\}$ and $\{a_p\}$ are defined in (\ref{eq:dddd}).

We denote here by $\Cov^{OU}$ the covariance matrix of the target OU bridge; the Karhunen-Lo\`eve expansion (\ref{eq:KL})
implies the eigenstructure $\{\lambda_{p,OU},\phi_p\}_{p=1}^{\infty}$ for $\Cov^{OU}$ with eigen-values:
\begin{equation*}
 \lambda_{p,OU} = \frac{1}{\frac{\pi^2\,p^2}{\ell^2}+\kappa^2}\ .
\end{equation*}
Examining the eigenvalues $\{P_p\}_{p=1}^{\infty}$ of $\mathcal{P}$, after some calculations one can verify that:
\begin{equation*}
\mathcal{P} \equiv \Cov_{OU}^{-1}\,\Cov \, \tfrac{\kappa^2\,h^2}{(2+\frac{h^2}{2})(1+\rho)} \ ,
\end{equation*}
so that:
\begin{equation}
\label{eq:pi}
  0\le P_p \le M\, \lambda_{p,OU}^{-1}\,\lambda_p\,\frac{1}{\ell^2} \ ,
\end{equation}
for some constant $M>0$.
Again, after some tedious calculations we get that:
\begin{equation}
\label{eq:ai}
 a_p = \lambda_p^{-1/2}\lambda_{p,OU}^{1/2}\,c_p\ ;\quad c_p^{2} :=\,\big(\,\frac{1}{c^2\,\kappa^2}-
\frac{1}{4\,p^2\pi^2}\,\big)^{-1} .
\end{equation}
Note that the term on the R.H.S.\@  of this last definition of $c_p^2$ is guaranteed to be positive due to
condition~(\ref{eq:condition}). In particular we have that:
\begin{equation}
\label{eq:ci}
 c_p^2 \le M\ ,
\end{equation}
for a constant $M>0$.
Now, starting from (\ref{eq:energy1}) we have that:
\begin{align}
\nonumber
\Exp\,[\,(\Delta H)^2\,] = \Exp\,[\,\langle \mathcal{A}x_0, x_0 \rangle^2 \,] +
\Exp\,[\,\langle \mathcal{B}v_0, &v_0 \rangle^2 \,] +
\Exp\,[\,\langle \mathcal{G}x_0, v_0 \rangle^2 \,] \\ &+ 2\,\Exp\,[\,\langle \mathcal{A}x_0, x_0 \rangle \,]\,
\Exp\,[\,\langle \mathcal{B}v_0, v_0 \rangle \,]
\label{eq:bigg}
\end{align}
as the rest of the expectations will be zero (this follows from the independency of the zero-mean variables $x_0, v_0$).
Henceforth, $\{A_p\}_{p=1}^{\infty}$, $\{B_p\}_{p=1}^{\infty}$, $\{G_p\}_{p=1}^{\infty}$ denote the eigenvalues of the operators $\mathcal{A}$, $\mathcal{B}$, $\mathcal{G}$ respectively.
Recalling that
$\langle \mathcal{A}x_0, x_0 \rangle = \sum_{p=1}^{\infty} A_p x_{0,p}^{2}$, we have:
\begin{align*}
\Exp\,[\,\langle \mathcal{A}x_0, & x_0 \rangle^2 \,] = \mathrm{Var}\,[\,\langle \mathcal{A}x_0, x_0 \rangle\,] +
\Exp^2\,[\,\langle \mathcal{A}x_0, x_0 \rangle \,] \\
&= 2\,\sum_{p=1}^{\infty} A_p^2\,\lambda_{p,OU}^{2} +
\big(\,\sum_{p=1}^{\infty} A_p\, \lambda_{p,OU}\,\big)^2\ .
\end{align*}
Doing similar calculations also for the rest of the terms on the R.H.S.\@ of (\ref{eq:bigg}), we have:
\begin{align*}
 \Exp\,[\,(\Delta H)^2\,] = 2\,\sum_{p=1}^{\infty} A_p^2\,\lambda_{p,OU}^{2} +
  2\,&\sum_{p=1}^{\infty} B_p^2\,\lambda_{p}^{2} +
\Big(\,\sum_{p=1}^{\infty}\big(\, A_p\,\lambda_{p,OU} + B_p\,\lambda_{p}  \,\big)\,\Big)^2\ \\
&+ \sum_{p=1}^{\infty} G_p^2\,\lambda_{p,OU}\,\lambda_{p}\ .
\end{align*}
Thus, using the representations and bounds in (\ref{eq:define}), (\ref{eq:pi}), (\ref{eq:ai}) and (\ref{eq:ci}), we have that:
\begin{equation*}
\sum_{p=1}^{\infty} A_p^2\,\lambda_{p,OU}^{2} \le M\, \sum_{p=1}^{\infty}\,\lambda_{p,OU}^{-2}\,\lambda_p^2\,
\frac{1}{\ell^4}\,\lambda_{p,OU}^2\,
= M\, \sum_{p=1}^{\infty}\frac{1}{p^4\,\pi^4} < \infty \ .
\end{equation*}
Similar calculations will give:
\begin{equation*}
\sum_{p=1}^{\infty} B_p^2\,\lambda_{p}^{2} \le M\, \sum_{p=1}^{\infty} \lambda_p^{-2}\,\lambda_{p,OU}^2\,c_p^4\,
\lambda_{p,OU}^{-2}\,\lambda_p^2 \,\frac{1}{\ell^4}\,\lambda_p^2
\le M\, \sum_{p=1}^{\infty}\frac{1}{p^4\,\pi^4} < \infty \ .
\end{equation*}
and:
\begin{equation*}
 \sum_{p=1}^{\infty} G_p^2\,\lambda_{p,OU}\,\lambda_{p}
\le M\, \sum_{p=1}^{\infty}\,\lambda_p^{-1}\,\lambda_{p,OU}\,c_p^2\,\lambda_{p,OU}^{-2}\,\lambda_p^2 \,\frac{1}{\ell^4}\,\lambda_{p,OU}\,\lambda_{p} \le M\, \sum_{p=1}^{\infty}\frac{1}{p^4\,\pi^4} < \infty \ .
\end{equation*}
Finally, we have that:
\begin{align*}
\big|\sum_{p=1}^{\infty}\big(\, A_p\,\lambda_{p,OU} + B_p\,\lambda_{p}  \,\big)\,\big| &=
\tfrac{1}{2}\,\big|\,\sum_{p=1}^{\infty}\sin^{2}(I\theta_p)\,P_p\,(-\lambda_{p,OU}+a_p^2\,\lambda_p)\,\big|\\
&= \tfrac{1}{2}\,\big|\,\sum_{p=1}^{\infty}\sin^{2}(I\theta_p)\,P_p\,\lambda_{p,OU}\,(c_p^2-1)\,\big|\\
& \le M\,\sum_{p=1}^{\infty}\,\lambda_p\,\frac{1}{\ell^2} = M\,\sum_{p=1}^{\infty} \frac{1}{p^2\pi^2} <\infty\ .
\end{align*}
Thus, we have proven that $\sup_l\,\Exp\,[\,\Delta H^2\,]<\infty$, which, as illustrated in the proof of Proposition \ref{prop:1}, is sufficient for our proof.

\end{proof}

\begin{proof}[Proof of Proposition \ref{prop:3}]
We will do our calculations using the right-most expressions in
(\ref{eq:bm}), (\ref{eq:bb}), where $\Cov^{bm}$, $\Cov^{bb}$ are respectively
specified.
For the first term in the expression for $\Phi$,
namely $\alpha = \alpha(x(u_1),x(u_2),\ldots, x(u_M))$, we can formally write
(with a slight abuse of notation,
in the equation that follows $\alpha$ denotes the mapping $x\mapsto \alpha(x(u_1),x(u_2),\ldots, x(u_M))$, as
it is the derivative of this path-mapping that we need to calculate):
\begin{equation*}
(\delta \alpha)(s) = \sum_{m=1}^{M}\alpha_m\cdot  \delta_{u_m}(s) \ ,
\end{equation*}
where $\delta_{u_i}$ is the Dirac delta function centered at $u_i$. Applying $\Cov^{bm}$ and $\Cov^{bb}$ will give immediately the
terms in the first lines of the expression for  $\Cov^{bm}\,\delta \Phi(x)$ and $\Cov^{bb}\,\delta \Phi(x)$
in the statement of the proposition.
For the second term $\beta = \beta(I_1,I_2,\ldots,I_L)$:
\begin{equation*}
(\delta \beta)(s) = \sum_{l=1}^{L}\beta_l\cdot z'_l(s,x(s)) \ .
\end{equation*}
Again, applying $\Cov^{bm}$ and $\Cov^{bb}$ will give the
terms in the second lines of the expression for  $\Cov^{bm}\,\delta \Phi(x)$ and $\Cov^{bm}\,\delta \Phi(x)$ in the statement of the proposition.

We proceed to the term $\gamma = \gamma(S_1,S_2,\ldots, S_J)$ with the stochastic integrals.
The algorithm applied in practice will involve a finite-difference approximation of the
stochastic integrals $\{S_j\}$.
 Below we will sacrifice accuracy at the notation to avoid taking too much
space for what involves otherwise straightforward derivative calculations.
Consider the discretised time instances $0=s_0<s_1<\cdots s_{N-1}<s_N = \ell$.
Denoting three consecutive discrete time instances among the  above by $s_{-}<s<s_{+}$, the finite-difference approximation, say $S_j^{N}$, of $S_j$
can be written as follows:
\begin{equation*}
S_j ^{N} =  \sum_{s\in\{s_1,\ldots,s_N\}} r_j(s_{-}, x(s_{-}))(x(s)-x(s_{-})) \ .
\end{equation*}
We can now calculate the partial derivative of $S_j^{N}$ w.r.t.\@
to the one of the $N$ variables, $x(s)$, making up the discretised path.
Notice that $x(s)$ will appear in two terms in the summation,
unless it is the last point $x(s_{N})$ of the $x$-vector when it will only appear once.
This explains the following calculation of the partial derivatives:
\begin{equation}
\label{eq:f1}
\frac{\partial S_j^{N}}{\partial x(s)} = \Delta q_j(s)\ ,\,\,\, s\in\{s_1,\ldots, s_{N-1}\} \ ; \,\,\,
\frac{\partial S_j^{N}}{\partial x(s_N)} = r_j(s_{N-1},x(s_{N-1}))\ ,
\end{equation}
where we have defined:
\begin{equation*}
\Delta q_j(s) = r{'}_{\!\!j}(s,x(s))(x(s_{+})-x(s)) - (r_j(s,x(s))-r_j(s_{-},x(s_{-})))\ .
\end{equation*}
Overall, we have that:
\begin{equation}
\label{eq:f2}
\frac{\partial \gamma}{\partial x(s)} = \sum_{j=1}^{J}\gamma_j\cdot\,\frac{\partial S_j^{N}}{\partial x(s)} \ .
\end{equation}
Then, the $N\times N$ discretised covariance operator $C^{bm}=(\min\{s_i,s_k\})_{i,k}$,
corresponding to the covariance matrix of a standard Brownian motion at the time instances $s_1,s_2,\ldots,s_N$ (this is the discretised version of $\Cov^{bm}$),
can be easily shown to apply as follows to a finite-dimensional vector $f\in \mathbb{R}^{N}$:
\begin{equation}
\label{eq:covNN}
(C^{bm}f)_{u} =  s_u\cdot \big(  \sum_{i=1}^{N} f_i \big)  -\sum_{k=1}^{u-1}\big( \sum_{i=1}^{k} f_i   \big)\,\Delta s_{k+1} \ , \quad u=1,2,\ldots, N\ ,
\end{equation}
where $\Delta s_{k+1}=s_{k+1}-s_k$.
Combining (\ref{eq:f1}), (\ref{eq:f2}), (\ref{eq:covNN}) will give (with $\nabla$ denoting the gradient):
\begin{align*}
(&C^{bm}\nabla\gamma)_u = \\ &\sum_{j=1}^{J}\gamma_j\cdot\,\Big(\,
s_u\,\big(\, r_j(s_{N-1},x(s_{N-1})) + \sum_{i=1}^{N-1}\Delta q_j(s_i)\,\big) -
\sum_{k=1}^{u-1}\big( \sum_{i=1}^{k}\Delta q_j(s_i)\big) \Delta s_{k+1}
 \,\Big)\ .
\end{align*}
It is easy to see now that a finite-difference approximation
 of the term appearing in the third line of the expression for $\Cov^{bm}\delta \Phi(x)$
in the statement of the Proposition  \ref{prop:3} would coincide precicely with the above expression.
A similar approach for the Brownian bridge case, considering the discrete time instances $0=s_0<s_1<\cdots s_{N-1}<s_N < s_{N+1} = \ell$ ,
and the corresponding $N$-dimensional matrix  $C^{bb}$ represented as below:
\begin{equation}
\label{eq:covNNa}
(C^{bb}f)_{u} =  \tfrac{s_u}{\ell}\cdot \sum_{k=1}^{N}\big( \sum_{i=1}^{k} f_i
 \big)\,\Delta s_{k+1}  -\sum_{k=1}^{u-1}\big( \sum_{i=1}^{k} f_i   \big)\,\Delta s_{k+1} \ , \,\, u=1,\ldots, N\ ,
\end{equation}
where now:
\begin{equation*}
S_j ^{N} =  \sum_{s\in\{s_1,\ldots,s_{N+1}\}} r_j(s_{-}, x(s_{-}))(x(s)-x(s_{-}))\ ;\quad
\frac{\partial S_j^{N}}{\partial x(s_i)} = \Delta q_j(s_i)\ ,
\end{equation*}
for $\Delta q_j(s)$ as defined earlier and $1\le i\le N$. This will give the calculation:
\begin{equation*}
(C^{bb}\,\nabla \gamma)_u = \sum_{j=1}^{J}\gamma_j\cdot\,\Big(\,
\frac{s_u}{\ell}\,\big(\, \sum_{k=1}^{N}\big( \sum_{i=1}^{k}\Delta q_j(s_i)\big) \Delta s_{k+1} -
\sum_{k=1}^{u-1}\big( \sum_{i=1}^{k}\Delta q_j(s_i)\big) \Delta s_{k+1}
 \,\Big)  \ ,
\end{equation*}
immediately recognised as the finite-difference discretisation of the term appearing in the third line of the expression for $\Cov^{bb}\delta \Phi(x)$
in the statement of the proposition.
%
%
%

%
\end{proof}

\begin{proof}[Proof of Proposition \ref{prop:transform}]
Consider a collection of discrete  time instances $0 <s_1 < s_2< \cdots < s_N$ with
$s_0=0$ and $s_N=\ell$  that include the data instances:
\begin{equation*}
\{u_1, u_2,\ldots, u_M\} \subset \{s_1, s_2,\ldots, s_N\} \ .
\end{equation*}
Let $\Delta s_i = s_i - s_{i-1}$. We will consider the following
finite-difference approximation $\Phi(x)=\Phi(x_1,x_2,\ldots,x_N)$ of the negative log-density:
\begin{align}
\Phi(x) &= \Phi_1(x)  +\Phi_2(x)  =  \alpha( v_{i_1},v_{i_2},\ldots, v_{i_M}  ) \label{eq:phiN} \\ +\,\,
&\beta\Big(\sum_{i=1}^{N}z_1(s_{i-1},v_{i-1})\Delta s_{i}, \sum_{i=1}^{N}z_2(s_{i-1},v_{i-1})\Delta s_{i}, \ldots,  \sum_{i=1}^{N}z_L(s_{i-1},v_{i-1})\Delta s_{i}
\Big)\nonumber
\end{align}
for indices $i_1, i_2, \ldots i_M$ such that $s_{i_m} = u_m$,
for $m=1,2,\ldots M$, and vector $v$ constructed via the finite-difference approximation:
\begin{equation*}
v_{i} = v_{i-1} + \mu(v_{i-1})\Delta s_i + \sigma(v_{i-1})\Delta x_{i} \ ,
\end{equation*}
for $i=1,2,\ldots, N$ with $v_0$ equal to a specified fixed initial condition.
We will be using the obtained expression in (\ref{eq:covNN}) for the $N\times N$ covariance matrix $C= C^{bm}=(\min\{s_i,s_j\})_{i,j}$  of a standard brownian motion at the time instances $s_1,s_2,\ldots,s_N$.
%
%
The function $\Phi:\mathbb{R}^{d}\mapsto \mathbb{R}$ in (\ref{eq:phiN}) and the matrix $C$ fully specify
a finite-difference approximation of the original target defined on the Hilbert space.

Now, we have the following recursion for the derivatives $$Y_{i,j}= \frac{\partial v_i}{\partial x_j}\ ,$$ for any $j\ge 1$:
\begin{align*}
 Y_{i,j} &= Y_{i-1,j} + \mu'(v_{i-1})\,Y_{i-1,j}\Delta s_i  + \sigma'(v_{i-1})\,Y_{i-1,j}\, \Delta x_i \ ;\quad  i>j+1 \\
 Y_{j+1,j} &= Y_{j,j} + \mu'(v_{j})\,Y_{j,j}\Delta s_{j+1}  + \sigma'(v_{i-1})\,Y_{j,j}\, \Delta x_{j+1} - \sigma(v_j) \ ;\\
 Y_{j,j} & = \sigma(v_{j-1})\\
 Y_{i,j} & = 0 \ ,\quad i < j \ .
\end{align*}
So, we can obtain that, for $i > j+1$:
\begin{equation*}
 \log( Y_{i,j} ) = \log(Y_{i-1,j}) + \log\big( 1 + \mu'(v_{i-1})\Delta s_i + \sigma'(v_{i-1})\Delta x_i \big)\ ,
\end{equation*}
and using this recursion we get that:
\begin{align}
\label{eq:Yi}
Y_{i,j}  &= \Delta Q_j \times e^{\sum_{k=j+2}^{i}\log\big(  1 + \mu'(v_{k-1})\Delta s_k + \sigma'(v_{k-1})\Delta x_k \big) } \ , \quad i \ge j+1\ ,\\
Y_{j,j}  &=  \sigma(v_{j-1}) \label{eq:YY}\\
Y_{i,j}  &= 0\ ,\quad i < j \ ,
\end{align}
where we have set:
\begin{equation*}
\label{eq:Y0}
\Delta Q_j \equiv Y_{j+1,j} = \sigma(v_{j-1})( \mu'(v_{j})\Delta s_{j+1} + \sigma'(v_{j})\Delta x_{j+1} )  - \Delta \sigma(v_j)\ ,
\end{equation*}
and $\Delta \sigma(v_j) = \sigma(v_j) -\sigma(v_{j-1})$.  We will also define for $1\le m \le M$ and $1\le l\le L$:
\begin{align*}
F_{m,r} &= \sum_{j=1}^{r} Y_{i_m,j} \ ,\quad r < i_m \ ; \\
G_{l,r} &= \sum_{j=1}^{r} \big( \sum_{i\ge j+1}z_l'(s_i,v_i)\Delta s_{i+1} \big)\, Y_{i,j} \ ; \\
J_{l,r} &= \sum_{j=1}^{r}z_l'(s_j,v_j)Y_{j,j}\,\Delta s_{j+1}\  .
\end{align*}
The above sequences will appear at the calculation of partial derivatives of $\Phi(x)$. It
is important to notice here that these sequences indeed constitute a finite-difference approximation of their
continuous-time counterparts appearing at the statement of the proposition: to see that
one only need to look at the analytical  definition of $Y_{i,j}$ in equations (\ref{eq:Yi})-(\ref{eq:YY}), and realise
that the sum $\sum_{k}\log\big(  1 + \mu'(v_{k-1})\Delta s_k + \sigma'(v_{k-1})\Delta x_k \big)$ is essentially a finite-difference approximation
of $\int (\mu'(v({u}))du + \sigma'(v({u}))d x(u))$ as for $\epsilon\approx 0$ we have that $\log(1+\epsilon)\approx \epsilon$.

We can now proceed to the calculation of the partial derivatives of $\Phi$. Clearly:
\begin{equation*}
\frac{\partial \Phi}{\partial x_j}  = \frac{\partial \Phi_1}{\partial x_j} + \frac{\partial \Phi_2}{\partial x_j} = \sum_{i\ge j}\big(  \frac{\partial \Phi_1}{\partial v_i } \cdot Y_{i,j} +
\frac{\partial \Phi_2}{\partial v_i } \cdot Y_{i,j}  \big) \ .
\end{equation*}
We can easily get:
\begin{align*}
 \sum_{i\ge j}\frac{\partial \Phi_1}{\partial v_i } \cdot Y_{i,j} = \sum_{m=1}^{M} \alpha_{m} Y_{i_m,j}\ .
\end{align*}
Using (\ref{eq:covNN}), tedious, but otherwise straightforward calculation will give that, for vector index $1\le u\le N$:
\begin{equation}
\label{eq:der1}
(C\,\nabla \Phi_1(x))_u =  \sum_{m=1}^{M} \alpha_m \,\Big(   s_{u\wedge i_m}\,(  F_{m,i_m-1}  + Y_{i_m,i_m}) + \sum_{k=1}^{u\wedge i_m-1} F_{m,k}\,\Delta s_{k+1} \Big)\ .
\end{equation}
Proceeding to the second term, $\Phi_2(x)$, we have that:
\begin{align*}
 \sum_{i\ge j}\frac{\partial \Phi_1}{\partial v_i } \cdot Y_{i,j} = \sum_{l=1}^{L} \beta_{l}\,\sum_{i\ge j }z_l(s_i,v_i)\Delta s_{i+1}\,Y_{i,j}\ .
\end{align*}
We now now multiply with the covariance matrix $C$ to obtain, after some calculations,
for $1\le u \le N$:
\begin{align}
\label{eq:der2}
(C\,\nabla \Phi_2(x))_u =  \sum_{l=1}^{L} \beta_l\, \Big(\,
s_u\,(\,  G_{l,N} + J_{l,N}\,) - \sum_{k=1}^{u-1} (\, G_{l,k}
 + J_{l,k} \, \big)\,\Delta s_{k+1}\,
 \Big)\ .
\end{align}
Upon inspection, (\ref{eq:der1})-(\ref{eq:der2}) provide the proof of the statement of the proposition.

\end{proof}

\bibliographystyle{plain}
\bibliography{references}

\end{document}